\documentclass[useAMS,usenatbib]{mn2e}

\usepackage{hyperref}

\usepackage{amssymb,epsfig,color}
\usepackage{amsmath}
\usepackage{graphicx}
\usepackage[font=footnotesize,caption=false]{subfig}
\usepackage{multirow}
\usepackage{xspace}

\newcommand{\FF}{\vphantom{\vdots}}
\newcommand{\lan}{\langle}
\newcommand{\ran}{\rangle}

\newcommand{\abs}[1]{\vert #1 \vert}
\newcommand{\avg}[1]{\left\langle #1 \right\rangle}
\newcommand{\dg}{^{\circ}}

\title[Astronomical Receiver Modelling Using Scattering Matrices]{Astronomical Receiver Modelling Using Scattering Matrices}
\author[O. G. King et al.]{O.G. King$^{1,2}$\thanks{E-mail: ogk@astro.caltech.edu}, 
Michael$\,$E. Jones$^{2}$,  
C. Copley$^{2,3}$, 
R.$\,$J. Davis$^{4}$,
J.$\,$P. Leahy$^{4}$, 
J. Leech$^{2}$,
\newauthor 
S.$\,$J.$\,$C. Muchovej$^{1}$, 
T.$\,$J. Pearson$^{1}$, 
and
Angela$\,$C. Taylor$^{2}$ \\
$^{1}$California Institute of Technology, Pasadena CA 91125, USA\\
$^{2}$Sub-department of Astrophysics, University of Oxford, Denys Wilkinson Building, Keble Road,
Oxford, OX1 3RH, UK\\
$^{3}$SKA South Africa, Park Road, The Park, Pinelands, Cape Town, South Africa \\
$^{4}$Jodrell Bank Centre for Astrophysics, School of Physics \& Astronomy, The University of Manchester, \\Oxford Road, Manchester, M13 9PL, UK }

\begin{document}

\date{Accepted XXX. Received YYY; in original form ZZZ}

\pagerange{\pageref{firstpage}--\pageref{lastpage}} \pubyear{2014}

\maketitle

\label{firstpage}

\begin{abstract}
Proper modelling of astronomical receivers is vital: it describes the systematic errors in the raw data, guides the receiver design process, and assists data calibration. In this paper we describe a method of analytically modelling the full signal and noise behaviour of arbitrarily complex radio receivers. We use electrical scattering matrices to describe the signal behaviour of individual components in the receiver, and noise correlation matrices to describe their noise behaviour. These are combined to produce the full receiver model.
We apply this approach to a specified receiver architecture: a hybrid of a continous comparison radiometer and correlation polarimeter designed for the C-Band All-Sky Survey. 
We produce analytic descriptions of the receiver Mueller matrix and noise temperature, and discuss how imperfections in crucial components affect the raw data.
Many of the conclusions drawn are generally applicable to correlation polarimeters and continuous comparison radiometers.
\end{abstract}


\section{Introduction}

Astronomical receiver modelling has many purposes. Perhaps the most important is understanding the data produced by the receiver: we want to know how the raw data values produced by the receiver relate to the astronomical signal of interest. We may, for instance, wish to know how sensitive the receiver will be or what systematic errors will be present in the data.
If the receiver is simple it may be straightforward to describe the raw data, but radio receiver architectures are often complex and difficult to model accurately. One approach to the complexity is to calculate the receiver response by numerical simulation. A more powerful approach is to model the receiver behaviour analytically: this in principle allows us to describe exactly how particular instrumental parameters affect the data, with no confusion as to what a particular artifact in the output is caused by.


Most analytic and semi-analytic approaches to characterizing systematic effects in receivers have employed Jones matrices to describe receiver components and Mueller matrices to characterize the effects of receiver imperfections on the observed signal, e.g., \citet{2001PASP..113.1274H,Carretti:2001p59,Hu:2003fa,ODea:2007kf}. In this formulation the propagation of radiation through a receiver can be described by a $2 \times 2$ Jones matrix $\mathbf{J}$. The total instrument Jones matrix is found by cascading (multiplying) the Jones matrices of the components in the instrument. If the receiver is polarization sensitive, the effect of the instrument on the true polarization vector can be found by calculating the Mueller matrix that describes the instrument. While powerful, a shortcoming of this approach is that it does not include the effect of noise produced by components in the receiver -- i.e., it does not model the sensitivity of the instrument.

In this paper we describe a method for modelling astronomical receivers that produces a full signal and noise description of the instrument.
Each component in a receiver is described by an electrical scattering matrix and a noise correlation matrix. These describe the signal response and noise properties of the component respectively. Scattering matrices provide a full description of the reflection and transmission of electromagnetic waves incident on a component; noise correlation matrices describe the noise produced by the component \citep{Zmuidzinas:2003fe,Pozar:2005}.
A component may have multiple ports (input and outputs); at radio and millimeter wavelengths these are easily understood as guided electromagnetic waves, while in a quasi-TEM optical system they might be thought of as different polarization states of the electromagnetic wave. 
We build a network of components by connecting their ports appropriately to describe the signal flow in the receiver. The response of the full network (the whole receiver) can then be calculated and its signal and noise response described by a single scattering matrix and noise correlation matrix respectively.

We will analyse a specific radio receiver architecture in this paper: the C-Band All-Sky Survey (C-BASS) receiver \citep{2014MNRAS.438.2426K}. This receiver measures the full polarization state of the instrument. It is a hybrid of two commonly used architectures -- a continuous comparison radiometer to measure the total intensity of the signal; and a correlation polarimeter to measure the linear polarization state of the signal. It can also measure the circular polarization state, though the architecture is not optimized to do this. Many of the results we obtain are applicable to correlation polarimeters and continuous comparison radiometers in general, rather than the specific C-BASS implementation described here.

In \S\ref{sec:stokes_parameters} we introduce the framework (Stokes parameters and Mueller matrices) we will use throughout the paper to describe the signal response of the receiver. In \S\ref{sec:receiver_modelling} we introduce the method, describing scattering matrices, noise correlation matrices, and how they can be used to derive a useful description of the instrument behaviour. In \S\ref{sec:C-BASS_receiver_model_and_errors} we describe the specific C-BASS receiver architecture. In \S\ref{sec:receiver_analysis} we analyse the receiver, produce exact descriptions of the instrument  signal and noise behaviour, and discuss how imperfections in crucial components affect the output data. In \S\ref{sec:testing_the_model} we test some of the predictions made in \S\ref{sec:receiver_analysis}, and fit some model parameters to the measured instrument response. We conclude in \S\ref{sec:conclusions}.

\section{Stokes Parameters} \label{sec:stokes_parameters}


The Stokes parameters are a convenient and powerful way of 
describing the state of polarization of an electromagnetic signal. $I$
describes the total intensity of the signal, $Q$ and $U$ describe the
linear polarization state, and $V$ describes the circular polarization
state. Stokes parameters are defined relative to a local coordinate
system; $Q$ represents the degree of linear polarization parallel and
perpendicular to the local coordinate axes, while $U$ represents the
linear polarization at 45 degrees to these
axes. Table~\ref{tab:stokes_parameters} lists the definitions of the
Stokes parameters in two commonly used bases of the electric field
vector: the orthogonal linear modes $E_{x}(t)$ and $E_{y}(t)$, and the
orthogonal circular modes $E_{l}(t)$ and $E_{r}(t)$.

The coherency vector \citep{Born:1964un} describes the state of an
electromagnetic signal by including all possible correlations between
its orthogonal electric field modes:
\begin{align}
\nonumber \mathbf{e} = & \avg{
\begin{bmatrix}
E_{x}(t)E_{x}^{*}(t) \\
E_{x}(t)E_{y}^{*}(t) \\
E_{y}(t)E_{x}^{*}(t) \\
E_{y}(t)E_{y}^{*}(t)
\end{bmatrix}
} \\
 = & \avg{ \mathbf{E}(t) \otimes \mathbf{E}^{*}(t)  },
\end{align}
where $\mathbf{E}(t)$ is the complex vector of the orthogonal linear electric field modes $E_{x}(t)$ and $E_{y}(t)$ of the signal, $\avg{ \ldots }$ indicates time averaging, and $\otimes$ indicates the Kronecker tensor product.

The Stokes vector [$I$, $Q$, $U$, $V$] is a
representation of the coherency vector in an abstract space. The
Stokes vector $\mathbf{e}^{S}$ is obtained from the coherency vector
$\mathbf{e}$ by
\begin{align}
 \mathbf{e}^{S} = & 
\begin{bmatrix}
I \\ Q \\ U \\ V
\end{bmatrix}
= \mathbf{Te} \label{eqn:Stokes_vector_from_coherency} \\
\textrm{where } \mathbf{T} = &
\begin{bmatrix}
 1 & 0 & 0 & 1 \\
 1 & 0 & 0 & -1 \\
 0 & 1 & 1 & 0 \\
 0 & -i & i & 0
\end{bmatrix}. \label{eqn:T_matrix_definition}
\end{align}
We see that $\mathbf{T}$ is a coordinate transformation of the
coherency vector to the abstract Stokes frame \citep{Hamaker:1996p442}.

\begin{table}
\caption{The Stokes parameters written in terms of orthogonal linear
  and orthogonal circular bases of the electric field vector. $\beta =
  4Rk_{B}$ is a proportionality constant to place the Stokes
  parameters in units of antenna temperature
  (Appendix~\ref{sec:stokes_in_temp_units}). The circular polarization bases are related to the linear polarization bases by the equations $E_l = \left(E_x+iE_y \right)/\sqrt{2}$ and $E_r = \left(E_x-iE_y \right)/\sqrt{2}$. $i$ is the imaginary number, while $\Re\{x\}$ and $\Im\{x\}$ are the real and imaginary parts of $x$ respectively.
}
\label{tab:stokes_parameters}
\begin{tiny}
\centering
\begin{tabular}{| c | c | c |}
\hline
 & Linear basis & Circular basis \\ \hline
$\beta I$ & $\avg{ \abs{  E_{x}(t) }^{2} } + \avg{ \abs{ E_{y}(t) }^{2} }$ & $\avg{ \abs{ E_{l}(t) }^{2} } + \avg{ \abs{ E_{r}(t) }^{2} }$ \\ \hline
\multirow{2}{*}{$\beta Q$} & \multirow{2}{*}{$ \avg{ \abs{ E_{x}(t) }^{2} } - \avg {\abs{ E_{y}(t) }^{2} }$} & $2\avg{ \Re\{E_{r}(t) E_{l}^{*}(t) \} }$ \\ 
 & &$\avg{E_l^*(t)E_r(t)+E_l(t)E_r^*(t)}$ \\ \hline
\multirow{2}{*}{$ \beta U$} & $2\avg{ \Re\{ E_{x}(t)E_{y}^{*}(t) \} }$  & $-2\avg{ \Im\{E_{r}(t) E_{l}^{*}(t) \} }$ \\ 
 & $\avg{E_x^*(t)E_y(t)+E_x(t)E_y^*(t)}$ & $i\avg{E_l^*(t)E_r(t)-E_l(t)E_r^*(t)}$ \\ \hline
\multirow{2}{*}{$\beta V$} & $ 2\avg{ \Im\{ E_{x}(t)E_{y}^{*}(t) \} }$ & \multirow{2}{*}{$\avg{ \abs{ E_{l}(t) }^{2} } - \avg{ \abs{ E_{r}(t) }^{2} } $} \\
 & $i\avg{E_x^*(t)E_y(t)-E_x(t)E_y^*(t)}$ & \\ \hline
\end{tabular}
\end{tiny}
\end{table}

The action of an optical element, or indeed the entire instrument, on
the Stokes vector of the astronomical signal can be represented by a
Mueller matrix. Suppose that the incident Stokes vector is given by
$\mathbf{e}^{S}$, and the Stokes vector of the signal after it has passed through an
optical element is $\mathbf{e}^{S}_{m}$. The Mueller matrix $\mathbf{M}$ that describes the optical element is then defined as
\begin{align}
  \mathbf{e}^{S}_{m} = & \mathbf{M} \mathbf{e}^{S}.
\end{align}

The elements of the Mueller matrix are given by:
\begin{align}
 \mathbf{M} = &
\begin{bmatrix}
 M_{II} & M_{IQ} & M_{IU} & M_{IV} \\
 M_{QI} & M_{QQ} & M_{QU} & M_{QV} \\
 M_{UI} & M_{UQ} & M_{UU} & M_{UV} \\
 M_{VI} & M_{VQ} & M_{VU} & M_{VV}
\end{bmatrix}.
\end{align}
The diagonal elements of the Mueller matrix encode the sensitivity to
each Stokes parameter, while the off-diagonal elements encode the
leakage between Stokes parameters.
If the instrument Mueller matrix elements are constant and measureable the Mueller matrix can be inverted and applied to the data to return a leakage-free data stream.

\section{Receiver Modelling Using Scattering and Noise Matrices} \label{sec:receiver_modelling}

\subsection{Scattering Matrix Modelling}

In the scattering matrix formulation any arbitrary component or network of components (excluding detectors) can be described by a frequency-dependent scattering matrix $\mathbf{S}(\nu)$. We omit the frequency dependence from now on for brevity.

The scattering matrix relates the incident, reflected, and transmitted waves that travel on transmission lines attached to the $N$~ports of a linear network. It provides a complete description of an $N$-port network as seen at its $N$~ports \citep{Pozar:2005}. This formulation can be extended to optical systems and used to describe instruments that contain both optical components, such as lenses or mirrors, and microwave circuit techniques, such as horns, transmission lines, filters, etc. \citep{Zmuidzinas:2003fe}.

\subsubsection{Scattering and noise matrices}

Consider an arbitrary $N$-port network. We denote the incident wave at port $i$ by $V_{i}^{+}$, the reflected wave by $V_{i}^{-}$, and the noise wave produced by the network at that port by $c_{i}$. These quantities are related by the scattering matrix $\mathbf{S}$ and noise wave vector $\mathbf{c}$ as:
\begin{equation}
\renewcommand{\arraystretch}{1.4}
\begin{bmatrix} 
V_{1}^{-}  \\  
\FF V_{2}^{-}  \\ 
\FF \vdots  \\ 
\FF V_{N}^{-}
\end{bmatrix}
= 
\begin{bmatrix} 
S_{11} & S_{12} & \cdots & S_{1N}  \\  
\FF S_{21} & 		       &  	      & \vdots         \\ 
\FF \vdots        &   		       &  	      &            \\ 
\FF S_{N1} & \cdots        &  	      & S_{NN} 
\end{bmatrix}
\begin{bmatrix} 
V_{1}^{+}  \\  
\FF V_{2}^{+}  \\ 
\FF \vdots  \\ 
\FF V_{N}^{+}
\end{bmatrix}
+
\begin{bmatrix} 
c_{1}  \\  
\FF c_{2}  \\ 
\FF \vdots  \\ 
\FF c_{N}
\end{bmatrix}
\label{eqn:scattering_matrix_description}
\end{equation}
The scattering matrix $\mathbf{S}$ is unitary if the device is lossless and reciprocal networks have symmetric scattering matrices \citep{Pozar:2005}.

The noise wave voltages $c_i$ of an $N$-port network are complex time-varying random variables characterized by a correlation matrix $\mathbf{C}$
\begin{align}
\nonumber \mathbf{C} = & \lan \mathbf{c} \otimes \mathbf{c}^{\dagger} \ran \\
= &
\begin{bmatrix} 
\lan \vert c_1 \vert^{2} \ran  & \lan c_1 c_2^* \ran  & \cdots & \lan c_1 c_N^* \ran  \\  
\FF \lan c_2 c_1^* \ran & 		       &  	      & \vdots         \\ 
\FF \vdots        &   		       &  	      &            \\ 
\FF \lan c_N c_1^* \ran & \cdots        &  	      & \lan \vert c_N \vert^{2} \ran
\end{bmatrix}
\end{align}
where $\dagger$ indicates the conjugate transpose operation, and $\mathbf{c}$ is a vector with elements $c_i$. The diagonal terms of $\mathbf{C}$ give the noise power deliverable at each port per unit bandwidth. The off-diagonal terms are correlation products. The noise correlation matrix $\mathbf{C}$ for a passive network is determined by its scattering matrix $\mathbf{S}$ \citep{Wedge:1991}
\begin{equation}
 \mathbf{C} = kT(\mathbf{I}-\mathbf{S}\mathbf{S}^{\dagger}) \label{eqn:noise_correlation_matrix_for_passive_device}
\end{equation}
where $k$ is Boltzmann's constant, $T$ is the physical temperature of the network, and $\mathbf{I}$ is the identity matrix. The noise correlation matrix for an active network can be determined by measurement or modelling.

\subsubsection{Solving the network response}

We can build a network of $N$-port devices, connected by nodes, and assign scattering matrices and noise correlation matrices to each device. Once all the components in a receiver have been described by scattering and noise correlation matrices, we can then calculate the scattering matrix and the noise wave vector that describe the whole receiver as seen at its inputs and outputs using the {\verb MATLAB }\footnote{\url{http://www.mathworks.com}} package {\verb SNS }\footnote{Download at \url{https://github.com/kingog/SNS}} \citep{King:2010p4771}. It implements algorithms that solve for the network response and can operate on both analytic and numeric descriptions of scattering and noise correlation matrices. The \texttt{SUPERMIX} software package \citep{1999stt..conf..268W} can also solve for the noise and signal response of a microwave network, but only numerically: it cannot provide an analytic description of the outputs.

\subsubsection{Interpreting the Measured Power} \label{sec:rewriting_in_stokes_and_noise}

\begin{figure}
 \centering
 \includegraphics[width=2.5in]{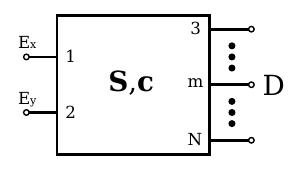}
\caption{A arbitrary receiver, where orthogonal linear polarization voltages $E_{x}(t)$ and $E_{y}(t)$ are presented at ports 1 and 2 respectively, while ports 3 to $N$ are the output ports. $D$ is the output at port $m$, and is connected to a power detector. The receiver is described by scattering matrix $\mathbf{S}$ and noise wave vector $\mathbf{c}$.}
 \label{fig:rd:mueller_matrix_from_scattering_matrix}
\end{figure}

We now have a scattering matrix and a noise wave vector that describe an arbitrary receiver architecture, excluding power detectors. The inputs to the scattering matrix are electric field vector elements from the sky, and the noise wave vector describes the noise at each output of the receiver. The outputs are connected to power detectors -- often square-law diodes in radio receivers. We now describe how to rewrite the power detected at each output in terms of Mueller matrices.

Consider the arbitrary receiver shown in Figure~\ref{fig:rd:mueller_matrix_from_scattering_matrix}. Orthogonal linear polarizations $E_{x}(t)$ and $E_{y}(t)$, representing either signals in transmission lines, orthogonal electric field modes in waveguide, or orthogonal electric field modes in free space, are connected to ports 1 and 2 of the receiver respectively. Receiver output $D$ at port $m$ is connected to a power detector. The receiver is described by the scattering matrix $\mathbf{S}$ and the noise wave vector $\mathbf{c}$. We will show how to rewrite the power per unit bandwidth detected at output $D$ as
\begin{align}
P_{D} = & P_{D,S} + P_{D,N} \label{eqn:diode_output_sky_and_noise}
\end{align}
where $P_{D,S} = k_{B}\left( M_{DI}I + M_{DQ}Q+M_{DU}U+M_{DV}V \right) $ is the contribution of the Stokes parameters that describe the sky signal voltages $E_{x}(t),E_{y}(t)$ to the power detected at the diode and $P_{D,N}$ is the contribution of receiver noise.

The Mueller matrix elements and the sky signal Stokes parameters in Equation~\ref{eqn:diode_output_sky_and_noise} are frequency dependent. The shape of the sky signal spectrum is required to obtain the band-integrated power.

\paragraph*{Stokes Contribution: }
The contribution $E_{m}(t)$ to output port $m$ from the input sky signal is given by
\begin{align}
E_m(t) = & S_{m1}E_{x}(t) + S_{m2}E_{y}(t).
\end{align}
The power contained in the signal $E_{m}(t)$ is then measured. At radio wavelengths this might be achieved through the use of a square-law detector diode. The measured power $P_{D,S}$ is given by:
\begin{align}
\nonumber P_{D,S} = & \alpha \lan E_{m}(t)E_{m}(t)^{*}\ran \\
\nonumber  = & \alpha\Big[\lan \vert E_{x}(t) \vert^{2}\ran \vert S_{m1}\vert^{2} + \lan \vert E_{y}(t) \vert^{2}\ran \vert S_{m2}\vert^{2} \\
\nonumber &+ \lan E_{x}(t)E_{y}^{*}(t) \ran S_{m1}S_{m2}^{*} \\
& + \lan E_{x}^{*}(t)E_{y}(t) \ran S_{m1}^{*}S_{m2} \Big] \label{eqn:rd:P_output_scattering}
\end{align}
where $\alpha = \alpha_{D}/(4R)$ scales the voltage squared units to power per unit bandwidth (see the Nyquist theorem Equation~\ref{eqn:nyquist_theorem}) and contains a factor $\alpha_{D}$ dependent on the power detection method and post-detector gain. We assume that the instrument scattering matrix parameters are constant during the averaging time period. Now let
\begin{align}
\nonumber  P_{D,S} = & k_{B}(M_{DI}I + M_{DQ}Q + M_{DU}U + M_{DV}V )\\
\nonumber = & \frac{1}{4R} \Big( M_{DI}\lan\vert E_{x}(t) \vert^{2} + \vert E_{y}(t) \vert^{2}\ran \\
\nonumber & + M_{DQ}\lan\vert E_{x}(t) \vert^{2}-\vert E_{y}(t) \vert^{2}\ran \\
\nonumber & + M_{DU}\lan E_{x}(t)E_{y}^{*}(t) +  E_{x}^{*}(t)E_{y}(t) \ran \\
& - iM_{DV}\lan E_{x}(t)E_{y}^{*}(t) - E_{x}^{*}(t)E_{y}(t)  \ran \Big) \label{eqn:rd:P_output_Mueller}
\end{align}
where we have used the definition of the Stokes parameters in a linear basis given in Table~\ref{tab:stokes_parameters}.

By comparing Equations~\ref{eqn:rd:P_output_scattering} and \ref{eqn:rd:P_output_Mueller} we can obtain the contribution of each Stokes parameter to the power measured at output D in terms of the scattering matrix parameters:
\begin{align}
\nonumber M_{DI} = & \frac{\alpha_{D}}{2} \left[ \vert S_{m1}\vert^{2} + \vert S_{m2}\vert^{2}  \right] \\
\nonumber M_{DQ} = & \frac{\alpha_{D}}{2} \left[ \vert S_{m1}\vert^{2} - \vert S_{m2}\vert^{2}  \right] \\
\nonumber M_{DU} = & \frac{\alpha_{D}}{2} \left[ S_{m1}S_{m2}^{*} + S_{m1}^{*}S_{m2}  \right] \\
 M_{DV} = & \frac{i\alpha_{D}}{2} \left[ S_{m1}S_{m2}^{*} - S_{m1}^{*}S_{m2}   \right].  \label{eqn:Mueller_from_S}
\end{align}

\paragraph*{Noise Contribution: }
We now describe how to derive the power contributed to output $D$ by the receiver noise using the noise wave vector returned by the network solving algorithm. We also show how to rewrite it referenced to the input of the receiver, i.e., as a receiver noise temperature.

If the noise wave vector of the receiver is given by $\mathbf{c}$, then the noise power measured at the output $D$ (port $m$ of the scattering matrix) in a 1~Hz bandwidth is given by $P_{D,N} = \alpha \lan c_{m}c_{m}^{*}\ran$, where $c_{m}$ is the noise wave vector element corresponding to output $D$.

We decompose the noise power seen at output $D$ into the power contributed by each noisy component. Suppose that component $k$ (of $M$ total noisy components in the receiver) is specified by a scattering matrix $\mathbf{S}^{k}$ and a noise wave vector $\mathbf{c}^{k}$. $c_{m}$ is given by
\begin{align}
c_{m} = & \sum_{k=1}^{M} c^{k}, \textrm{where } c^{k} = \mathbf{b}^{k} \mathbf{c}^{k}.
\end{align}
$c^{k}$ is the weighted contribution of the elements of the noise wave vector $\mathbf{c}^{k}$  to the total noise wave signal seem at port $m$. 
$\mathbf{b}^{k}$ is a row vector containing the weights; it is some function of the receiver response and is calculated during the network solving step.

Noise waves from different devices are usually not correlated\footnote{However, common temperature fluctuations of the amplifiers can cause a correlated noise component.}: $\lan c_{i}^{k}(c_{j}^{p})^{*} \ran = 0$ for $k\neq p$. So, $P_{D,N}$ is given by the sum of the individual component contributions:
\begin{align}
\nonumber P_{D,N} = & \alpha \sum_{k=1}^{M}P_{D,N}^{k} \\
\textrm{where }P_{D,N}^{k} = &  \mathbf{C}^{k}\cdot \big(\mathbf{b}^{k} \otimes (\mathbf{b}^{k})^{\dagger}\big).
\end{align}
Here $\mathbf{C}^{k}$ is the noise correlation matrix for component $k$ and $\cdot$ is the matrix dot product.

We have rewritten the sky contribution $P_{D,S}$ to the detected power in terms of the Stokes parameters. If we want to rewrite the noise contribution in a way that we can directly compare to the Stokes contributions we can turn it into a receiver temperature by referencing it to the receiver input. The receiver temperature $T_{D}$ of output $D$ is defined as the temperature of a thermal source seen equally at each receiver input that, for a noiseless receiver, produces the same power $P_{D,N}$ at output $D$ as the noise does:
\begin{align}
\nonumber P_{D,N} = & \alpha \Big( \abs{S_{m1}}^{2} + \abs{S_{m2}}^{2} \Big) T_{D} \\
\therefore T_{D} = & \frac{\sum_{k=1}^{M} P_{D,N}^{k}}{\abs{S_{m1}}^{2} + \abs{S_{m2}}^{2} }.
\end{align}

\paragraph*{Noise variance:} The previous section described the power seen at a particular detector due to noise produced by components in the receiver. The variance of the power signal can be obtained from the radiometer's equation:
\begin{align}
 \sigma_{D} = \frac{P_{D,S}+P_{D,N}}{ \left( \abs{S_{m1}}^{2} + \abs{S_{m2}}^{2} \right) \sqrt{\Delta\nu \tau}}
\end{align}
where $\Delta\nu$ is the signal bandwidth and $\tau$ is the integration time. We have turned the detected power $P_{D,S}+P_{D,N}$ into an antenna temperature by referencing it to the input using the gain term $\abs{S_{m1}}^{2} + \abs{S_{m2}}^{2}$.

\section{C-BASS Receiver Model} \label{sec:C-BASS_receiver_model_and_errors}

We apply the scattering matrix modelling approach to the northern C-BASS receiver (described in \citealt{2014MNRAS.438.2426K}).

\begin{figure*}
 \centering
 \includegraphics[width=\textwidth]{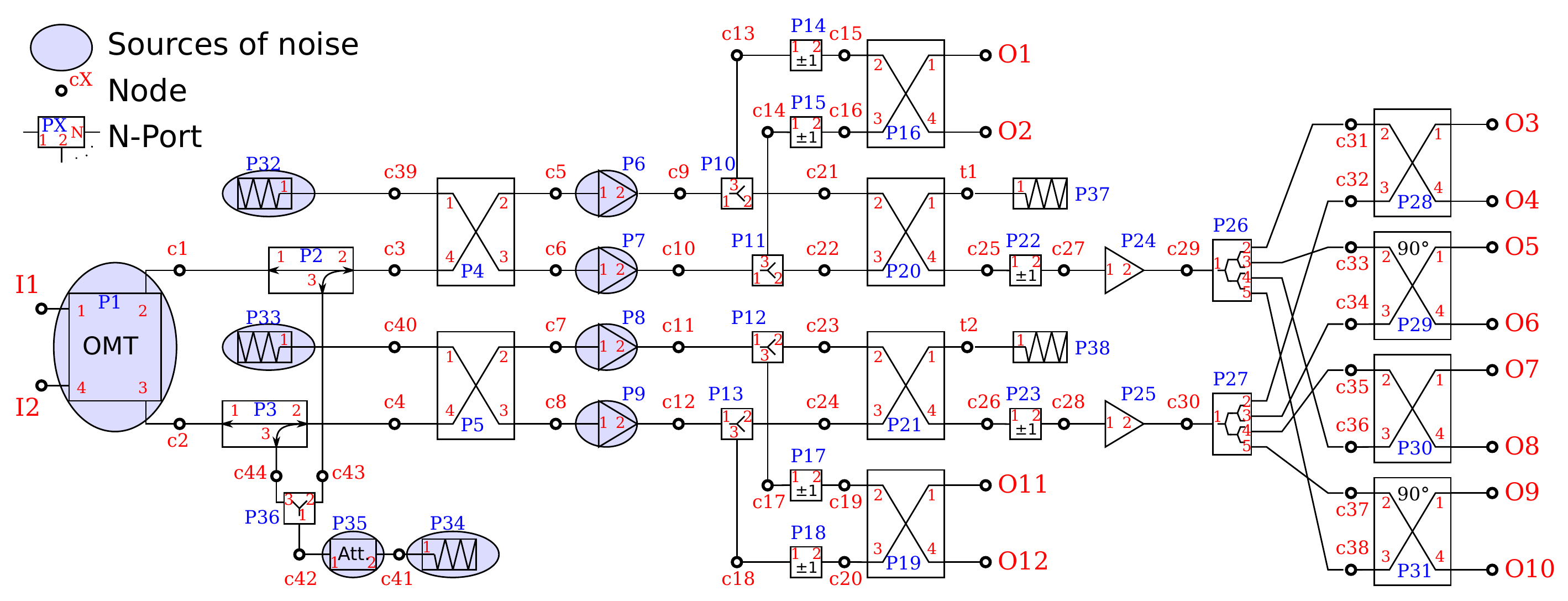}
\caption{A model of the C-BASS receiver used in the analysis. It is a hybrid of a continuous comparison radiometer and a correlation polarimeter.
Components are modelled as $N$-port devices (numbered P1 to
  P36) connected by central nodes (numbered c1 to c44).  The inputs to
  the receiver are orthogonal linear electric field vectors from the
  horn; $E_{x}(t)$ is connected to input node I1 and $E_{y}(t)$ is
  connected to input node I2. Output nodes O1 to O12 are connected to
  power detectors. The OMT, cold reference loads, calibration noise
  diode and attenuator, and amplifiers produce noise.}
 \label{fig:receiver_model}
\end{figure*}

The northern C-BASS receiver is a combination of a continuous comparison radiometer and a correlation polarimeter, shown in Figure~\ref{fig:receiver_model}. 
The radiometer measures the powers of both orthogonal circular polarizations independently by correlating them against two independent  stabilised thermal loads, so fluctuations
in the measured quantities as the telescope scans the sky track the true sky brightness. This continuous-comparison radiometer architecture reduces the $1/f$ gain
fluctuations, at the expense of a $\sqrt{2}$ higher white noise level due to the thermal noise of the load.

The linearly polarized Stokes parameters $Q$ and $U$ are measured simultaneously by a complex cross-correlation between the orthogonal
circular polarizations.  This is exactly the same process that is used in interferometric polarimeters, except that in this case the
orthogonally polarized signals are from the same antenna rather than two different antennas.  The $1/f$ gain fluctuations due to the low
noise amplifiers are uncorrelated between amplifiers and so are removed in the correlation operation.

The model of the receiver used in the systematic error analysis is
shown in Figure~\ref{fig:receiver_model}. Components are modelled as
$N$-port devices (labelled P1 to P36) connected by central nodes
(labelled c1 to c44).  The inputs to the receiver are orthogonal
linear electric field vectors from the horn; $E_{x}(t)$ is connected
to input node I1 and $E_{y}(t)$ is connected to input node I2. Output
nodes O1 to O12 are connected to power detectors. In this model we assume that the OMT, cold
reference loads, calibration noise diode and attenuator, and
amplifiers produce noise. Every component will produce noise, but these components will dominate the noise budget.

The model shown in Figure~\ref{fig:receiver_model} accurately
represents the action of the receiver, though it is not a facsimile of
the actual receiver diagram. Long gain chains, composed of multiple
amplifiers, attenuators, filters, isolators, and slope compensators
are represented by a single amplifier. Not all components are modelled
as producing noise, though all significant sources are modelled. A
fully-parameterized component-for-component reproduction of the real
receiver is not suitable for analytic description; the elements of the
Mueller matrix become so complex as to be meaningless and would fail
to illuminate the important lessons that can be learnt using a
simpler, but representative, model.

\subsection{Receiver Data Channels}

In \S\ref{sec:rewriting_in_stokes_and_noise} we described how to
rewrite the power detected at each receiver output in terms of
contributions from the sky (written in terms of Stokes parameters) and
contributions from noisy components in the receiver. This allows us to
create a description of each receiver data channel in a framework
that is powerful and natural to radio astronomy. We will 
express the vector $\mathbf{r}$ of receiver data channels (power per
unit bandwidth) in the form:
\begin{align}
\nonumber \mathbf{r} = & k_{B} \Big( \mathbf{M}_{\rm{corr}} \big( \mathbf{M}_{\rm{OMT}}\mathbf{e}^{S} + \mathbf{N}_{\rm{OMT}} \big) + \mathbf{N}_{\rm{corr}} \Big)  \\
\nonumber = & k_{B} \Big( \mathbf{M}_{\rm{rec}} \mathbf{e}^{S} + \mathbf{N}_{\rm{rec}}  \Big) \\
\nonumber  & \rm{where:} \\
\nonumber \mathbf{M}_{\rm{rec}} = & \mathbf{M}_{\rm{corr}} \mathbf{M}_{\rm{OMT}} \\
 \mathbf{N}_{\rm{rec}} = & \mathbf{M}_{\rm{corr}}\mathbf{N}_{\rm{OMT}}+\mathbf{N}_{\rm{corr}}. \label{eqn:receiver_data_vector_definition} 
\end{align}
Here we have split the receiver into two parts at the point where the noise diode calibration signal is injected. We refer to the pre-calibration signal injection part of the receiver as the OMT section, and the post-signal injection part as the correlator section. $\mathbf{M}_{\rm{OMT}}$ and $\mathbf{M}_{\rm{corr}}$ are the Mueller matrices for the OMT and correlator sections respectively. 
The contributions of the noise sources to each data stream are given by $\mathbf{N}_{\rm{OMT}}$ and $\mathbf{N}_{\rm{corr}}$, which describe the OMT section and correlator section (amplifiers, lossy components, calibration noise sources) respectively.
The action of the telescope optics (reflectors and horn) can be included by prepending its Mueller matrix to the Mueller matrix chain.


In this receiver outputs O1 to O12 are connected to detector diodes whose outputs are described by Equation~\ref{eqn:diode_output_sky_and_noise}. We subtract the signals from adjacent detectors to give us 6 data streams labelled I1 (O2$-$O1), I2 (O12$-$O11), U1 (O3$-$O4), Q1 (O6$-$O5), U2 (O7$-$O8), and Q2 (O9$-$O10). These can each be written in terms of Stokes parameter contributions and noise contributions.  We then calculate the raw receiver data streams:
\begin{align}
\mathbf{r}  =
 \begin{bmatrix}
 r_I \\
 r_Q \\
 r_U \\
 r_V
 \end{bmatrix} =
\begin{bmatrix}
\rm{I1}+\rm{I2} \\
(\rm{Q1}+\rm{Q2})/2 \\
(\rm{U1}+\rm{U2})/2 \\
\rm{I1}-\rm{I2}
\end{bmatrix}. \label{eqn:CBASS_receiver_data_vector}
\end{align}

\subsection{Analysis Procedure}

In the analysis that follows we construct a receiver model using {\verb SNS } and the {\verb MATLAB } symbolic algebra toolbox in which all the components have fully parameterized scattering matrices as described in \S\ref{sec:component_scattering_matrices}.  We perform the computationally expensive steps of calculating the analytic receiver scattering matrix and noise wave vector, and deriving the Mueller matrix $\mathbf{M}_{\rm{rec}}$ and noise vector $\mathbf{N}_{\rm{rec}}$, only once. The elements of $\mathbf{M}_{\rm{rec}}$ and $\mathbf{N}_{\rm{rec}}$ are algebraic expressions that contain a full description of the receiver. We explore the effect of imperfections in particular components on the instrument performance by removing unwanted parameters -- this is achieved by substituting ``perfect'' values for the error parameters (0 or 1, depending) and simplifying the resultant expressions.

\subsection{Component Models} \label{sec:component_scattering_matrices}

In general there are three levels of non-ideality in the scattering matrix that describes a component: those implicit in the design, random variations from device to device, and measurement errors. For the components in the C-BASS receiver the non-ideal behaviour implicit in the design was generally dominant: device to device variations were substantially lower and measurement error was negligible. We assume in this analysis that all errors are implicit to the design, and hence we can describe nominally identical components with the same matrices.

A comment on notation: amplitude balance errors are denoted by the $\delta$ symbol, and are zero for an ideal component. Phase errors are denoted by $\phi$ (zero for an ideal component), and transmission amplitudes are denoted by $\alpha$ (one for an ideal component). This convention was adopted to make it easy to verify that the derived Mueller matrices were sensible -- a quick visual substitution for ideal component values should result in the identity matrix.

\subsubsection{Circularizing OMT}

 We model the circularizing OMT (P1) as a component that accepts as
 inputs the orthogonal linear components of the electric field from
 the sky and produces at its outputs orthogonal circularly polarized
 signals. In the C-BASS receiver this is achieved by first extracting
 orthogonal linear $\rm{TE}_{11}$ modes from a circular waveguide
 using 4 rectangular probes
 \citep{2014MNRAS.438.2426K,Grimes:2006ef}. The out-of-phase signals
 from opposite pairs of probes are then combined with a $180^{\circ}$
 phase shift, using two $180^{\circ}$ hybrids, to obtain two orthogonal
 linear polarizations. Finally, these linear polarizations are then
 passed through a $90^{\circ}$ hybrid to produce two orthogonal circular
 polarizations. In our case, the two $180^{\circ}$ hybrids and single
 $90^{\circ}$ hybrid were fabricated on a single planar substrate to
 form a device known as a linear-to-circular converter. For
 simplicity, we hereinafter refer to the combination of the 4-probe
 linear OMT and the linear-to-circular converter as the
 \emph{circularizing OMT}.

 The scattering matrix for an ideal circularizing OMT that accepts $E_{x}(t)$
at port 1 and $E_{y}(t)$ at port 4 and returns the circular
polarization signals at ports 2 and 3 is the same as that for an ideal
$90\dg$ hybrid:

\begin{align}
\rm{Ideal\ } \mathbf{S}_{\rm{o}} = & \frac{1}{\sqrt{2}}
\begin{bmatrix} 
0  & 1  & i & 0  \\  
1  & 0  & 0 & i  \\  
i  & 0  & 0 & 1  \\  
0  & i  & 1 & 0  
\end{bmatrix}.
\end{align}

\begin{figure}
 \centering
 \includegraphics[width=3in]{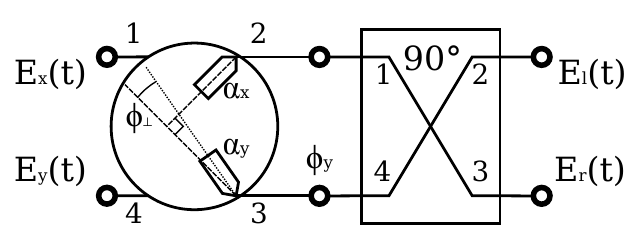}
\caption{Scattering matrix model of the circularizing OMT. Linear polarization signals are presented to a linear OMT. The $x$ axis of the OMT is perfectly aligned with the $x$ component of the sky signal, but the $y$ axis is rotated by $\phi_{\perp}$ from nominal, leading to leakage of $E_{x}$ into the $y$ output. Each probe has gain $\alpha_{x,y}$ (ideally 1), and there is a phase shift in the $y$ line $\phi_{y}$ relative to the $x$ line. The linear OMT is followed by a $90\dg$ hybrid to circularise the voltages.}
 \label{fig:omt_model}
\end{figure}

 Any real OMT will not perfectly convert linear polarizations into
 orthogonal circular polarizations. In practice, the non-ideal
 behaviour can arise due to mismatches in probe dimensions, probe
 angles or non-ideal performance of the $180^{\circ}$ and $90^{\circ}$
 hybrids which make up the linear-to-circular converter. We may model
 this non-ideal behaviour by referring to a somwehat simplified,
 conceptual representation of the circularizing OMT
 (Figure~\ref{fig:omt_model}). Linear polarized signals can be thought
 of as being presented to a linear OMT at ports 1 and 4. The $x$
 output (port 2) of the OMT is perfectly aligned with the $x$
 component of the sky signal, but the $y$ output (port 3) is rotated
 by $\phi_{\perp}$ from the nominal (perpendicular) orientation,
 leading to leakage of $E_{x}$ into the $y$ output. Each probe has a
 gain $\alpha_{x,y}$ (ideally 1), and there is a phase shift
 $\phi_{y}$ in the $y$ line relative to the $x$ line. This simplified
 linear OMT is then followed by a standard, imperfect, $90\dg$ hybrid
 (described by Equation~\ref{eqn:90hybrid_error_model}). The fully
 parameterized model of the imperfect circularizing OMT is then given
 by:

\begin{align}
\mathbf{S}_{\rm{o}} = & \frac{1}{\sqrt{2}}
\begin{bmatrix} 
0  & S_{lx}  & S_{rx}  & 0   \\  
S_{lx}  & 0  & 0  & S_{ly}  \\  
S_{rx}   & 0  & 0  & S_{ry}  \\  
0 & S_{ly} & S_{ry} & 0
\end{bmatrix} \label{eqn:scattering:OMT} \\
\nonumber S_{lx} = & \sqrt{\alpha_{x}}\sqrt{1+\delta}+i\sqrt{\alpha_{y}}\sqrt{1-\delta}\sin\phi_{\perp} e^{-i(\phi_{90}+\phi_{y})} \\
\nonumber S_{rx} = & i\sqrt{\alpha_{x}}\sqrt{1-\delta}e^{-i\phi_{90}}+\sqrt{\alpha_{y}}\sqrt{1+\delta}\sin\phi_{\perp} e^{-i\phi_{y}} \\
\nonumber S_{ly} = & i\sqrt{\alpha_{y}}\sqrt{1-\delta}\cos\phi_{\perp} e^{-i(\phi_{90}+\phi_{y})} \\
\nonumber S_{ry} = & \sqrt{\alpha_{y}}\sqrt{1+\delta}\cos\phi_{\perp} e^{-i\phi_{y}}.
\end{align}

As this is a passive component the noise correlation matrix can be
determined using
Equation~\ref{eqn:noise_correlation_matrix_for_passive_device}. We
emphasize that the parameters introduced in Equation
\ref{eqn:scattering:OMT} are simply describing the non-ideality of the
complete circularizing OMT, and will not necessarily correspond to
e.g., the physical probe angles within the four-probe OMT itself.

\subsubsection{$180\dg$ Hybrid}

The $180\dg$ hybrid (P4, P5, P16, P19, P20, P21, P28, and P30 in Figure~\ref{fig:receiver_model}) is a component that combines two incoming voltages, producing at one output port the sum of the inputs and at the other the difference of the inputs. It is sometimes called a ``magic tee'' when implemented in waveguide. The scattering matrix for an ideal $180\dg$ hybrid, with ports 1 and 4 being the inputs and ports 2 and 3 being the outputs, is given by:
\begin{align}
\rm{Ideal:\ } \mathbf{S}_{180} = 
\frac{1}{\sqrt{2}}
\begin{bmatrix} 
0  & 1  & 1 & 0  \\  
1  & 0  & 0 & -1  \\  
1  & 0  & 0 & 1  \\  
0  & -1  & 1 & 0  
\end{bmatrix}.
\end{align}

A well-designed hybrid can closely approximate the ideal behaviour, but the two most significant types of imperfect behaviour that will remain are amplitude and phase imbalances. Amplitude imbalances occur if the power from one input port is not equally split between the output ports. We use the parameter $\delta_{180}$ to designate this imbalance. Phase imbalances occur if there are phase errors in the ``sum'' and ``difference'' outputs: $\phi_{\Sigma}$ indicates the phase error in the summation operation and $\phi_{\Delta}$ indicates the phase error in the difference operation. The symmetric, unitary, scattering matrix for such a hybrid is given by:
\begin{align}
& \nonumber \rm{Error:\ } \mathbf{S}_{180} = \\
& \frac{1}{\sqrt{2}}
\begin{bmatrix} 
\FF 0  &\cdots  &  &   \\  
\FF \sqrt{1+\delta_{180}}  & 0  &  & \\  
\FF \sqrt{1-\delta_{180}}e^{-i\phi_{\Sigma}}  & 0  & 0 & \vdots   \\  
\FF 0  & -\sqrt{1-\delta_{180}}e^{-i\phi_{\Delta}}  & \sqrt{1+\delta_{180}}  & 0  
\end{bmatrix}. \label{eqn:180hybrid_error_model}
\end{align}
Note that this matrix is equal to the ideal matrix if $\delta_{180}=\phi_{\Sigma}=\phi_{\Delta}=0$.

The C-BASS receiver shown in Figure~\ref{fig:receiver_model} contains
several stages of $180\dg$ hybrids. Those in the first stage, between
the circularizing OMT and the first amplification stage, are called the cold hybrids as they are located in the receiver cryostat. Those in later stages are called warm hybrids.

\subsubsection{$90\dg$ Hybrid}

A $90\dg$ hybrid (P29 and P31) combines two input voltages with equal amplitude but with a $90\dg$ phase difference. They are commonly used to introduce $90\dg$ phase shifts to a voltage (in correlators, for instance), and to convert linear polarization signals to a circular basis in correlation polarimeters. The scattering matrix for an ideal $90\dg$ hybrid is given by:
\begin{align}
 \rm{Ideal:\ } \mathbf{S}_{90} =  
\frac{1}{\sqrt{2}}
\begin{bmatrix} 
0  & 1  & i & 0  \\  
1  & 0  & 0 & i  \\  
i  & 0  & 0 & 1  \\  
0  & i  & 1 & 0  
\end{bmatrix}.
\end{align}

As with the $180\dg$ hybrid, both amplitude and phase imbalances can occur in these hybrids. However, the C-BASS $90\dg$ hybrids were implemented as microstrip  branch-line couplers. These are symmetric structures that remove the need for having two different phase errors for the outputs as we do with the $180\dg$ hybrid. We describe the amplitude imbalance using the parameter $\delta_{90}$ and the phase imbalance using the parameter $\phi_{90}$. The scattering matrix for the $90\dg$ hybrid is then given by the symmetric, unitary, matrix:
\begin{align}
& \nonumber \rm{Error:\ } \mathbf{S}_{90} = \\
& \frac{1}{\sqrt{2}}
\begin{bmatrix} 
\FF 0  &\cdots  &  &   \\  
\FF \sqrt{1+\delta_{90}}  & 0  &  & \\  
\FF i\sqrt{1-\delta_{90}}e^{-i\phi_{90}}  & 0  & 0 & \vdots   \\  
\FF 0  & i\sqrt{1-\delta_{90}}e^{-i\phi_{90}}  & \sqrt{1+\delta_{90}}  & 0  
\end{bmatrix}. \label{eqn:90hybrid_error_model}
\end{align}

\subsubsection{Directional Coupler}

Directional couplers (P2 and P3) are used to inject a calibration signal from a broadband noise source into the signal path. This is used to calibrate the instrument by measuring the gain of the receiver and the polarization vector rotation angle of the receiver. The scattering matrix for an ideal directional coupler where the through signal is connected to port 1, the coupled signal is connected to port 3 and the output is port 2 is given by:
\begin{align}
\mathbf{S}_{\rm{cpl}} = &
\begin{bmatrix} 
0  & 1  & 0 \\  
1  & 0  & \sqrt{D} \\  
0  & \sqrt{D}  & 0
\end{bmatrix}.
\end{align}
Here $D$ is the power coupling factor. Technically, if the coupler is lossless the through-parameter should be $S_{21}=S_{12}=\sqrt{1-D}$, but if the coupling factor $D$ is sufficiently low (say $D=-30$~dB) it can be assumed to be equal to 1.

\subsubsection{Cold Reference Loads and Terminations}

The cold reference loads are modelled as perfectly-matched 1-port terminations. Their scattering matrices and noise correlation matrices are given by:
\begin{align}
\nonumber \mathbf{S}_{\rm{load}} = & \begin{bmatrix} 0 \end{bmatrix} \\
\mathbf{C}_{\rm{load}} = & kT_{\rm{load}}\begin{bmatrix} 1 \end{bmatrix}
\end{align}
where $T_{\rm{load}}$ is the physical temperature of the load, and $\begin{bmatrix} 1 \end{bmatrix}$ is a $1 \times 1$ unit matrix. The cold reference loads P32 and P33 have physical temperatures $T_{A}$ and $T_{B}$ respectively. Ideally $T_{A} = T_{B}$, but there may be some small remaining temperature difference. The terminations P37 and P38 contribute a negligible amount of noise so are assigned a 0~K physical temperature.

\subsubsection{Noise Diode}

The noise diode (P34) is modelled as a termination at physical temperature $T_{\rm ND}$. When the noise diode is turned off $T_{\rm ND}[\rm{off}] = 290$~K. When the noise diode is turned on the temperature changes to $T_{\rm ND}[\rm{on}] = 290(1+10^{{\rm ENR}/10})$, where ${\rm ENR}$ is the Excess Noise Ratio of the noise diode in dB.

\subsubsection{Attenuator}

The noise diode is followed by an attenuator (P35) to control the injected signal level. The scattering matrix and noise correlation matrix for a perfect attenuator are given by:
\begin{align}
\nonumber \mathbf{S}_{\rm{att}} = &
\begin{bmatrix}
0 & \sqrt{L} \\
\end{bmatrix}\\
\mathbf{C}_{\rm{att}} = & kT_{\rm{amb}}
\begin{bmatrix}
1-L & 0 \\
0 & 1-L
\end{bmatrix}
\end{align}
where $T_{\rm{amb}}$ is the temperature of the attenuator and $L$ is the power transmission.

\subsubsection{Amplifier}

An amplifier (P6, P7, P8, P9, P24, and P25) working in the linear regime (i.e., uncompressed) increases the amplitude of the incoming voltage wave by some factor. However, this comes at the expense of adding noise to the signal. The ideal scattering matrix and noise correlation matrix for such an amplifier where the input is port 1 and the output is port 2 is given by:
\begin{align}
\nonumber \mathbf{S}_{\rm{amp}} = &
\begin{bmatrix}
0 & 0 \\
g_{\rm{amp}} & 0
\end{bmatrix} \\
\mathbf{C}_{\rm{amp}} = &
\begin{bmatrix}
0 & 0 \\
0 & kT_{\rm{amp}}\abs{g_{\rm{amp}}}^{2}
\end{bmatrix}.
\end{align}
Here $g_{\rm{amp}}$ is the complex voltage gain of the amplifier. It has a power gain factor of $\abs{g_{\rm{amp}}}^{2}$ and introduces a phase shift of $\angle g_{\rm{amp}}$ to the wave. The noise power produced at the output port of the amplifier is given by $kT_{\rm{amp}}\abs{g_{\rm{amp}}}^{2}$, where $T_{\rm{amp}}$ is the noise temperature of the amplifier.

Note that in reality amplifiers produce noise at their input port as well and this noise is correlated with the output noise. If this noise were to leak through to other signal chains through the pre-amplifier components it would introduce a spurious correlation to the signal. The resulting offset in the polarization channels will be constant with time and telescope pointing. However, if the first-stage amplifiers are preceded by isolators that severely attenuate signals travelling in the reverse direction while allowing signals in the forward direction to pass nearly unattenuated,
this small but constant offset can be substantially reduced.

\subsubsection{Power Divider}

The receiver in Figure~\ref{fig:receiver_model} contains both 2-way (P10, P11, P12, P13, and P36) and 4-way (P26 and P27) power dividers. If port 1 is the input power then they have ideal scattering matrices given by:
\begin{align}
\mathbf{S}_{\textrm{2-way}} = & \frac{1}{\sqrt{2}}
\begin{bmatrix}
0 & 1 & 1 \\
1 & 0 & 0 \\
1 & 0 & 0
\end{bmatrix} \\
\mathbf{S}_{\textrm{4-way}} = & \frac{1}{2}
\begin{bmatrix}
0 & 1 & 1 & 1 & 1 \\
1 & 0 & 0 & 0 & 0 \\
1 & 0 & 0 & 0 & 0 \\
1 & 0 & 0 & 0 & 0 \\
1 & 0 & 0 & 0 & 0 
\end{bmatrix}.
\end{align}

\subsubsection{Phase Switch}

Phase switches (P14, P15, P17, P18, P22, and P23) are used to introduce a $0\dg$ or $180\dg$ phase shift to a signal. They remove the effects of post-detection gain variations, as demonstrated in \S\ref{sec:role_of_phase_switching}. The scattering matrix for an ideal phase switch is given by:
\begin{align}
\rm{Ideal:\ } \mathbf{S}_{\rm{ps}} = 
\begin{bmatrix}
 0 & \pm 1 \\
\pm 1 & 0
\end{bmatrix}.
\end{align}
The sign of the transmitted voltage is switched.

In a real phase switch we can have two sources of error: transmission amplitude differences between the two phase switch states, and a phase error (i.e., retard the phase by something other than $180\dg$). A more realistic time-dependent phase switch model is given by:
\begin{align}
\rm{Error:\ } \mathbf{S}_{\rm{ps}}(t) =
\begin{bmatrix}
0 & \sqrt{\alpha_{0,1}(t)}e^{-i\phi_{0,1}(t)} \\
 \sqrt{\alpha_{0,1}(t)}e^{-i\phi_{0,1}(t)} & 0
\end{bmatrix}. \label{eqn:scattering:phase_switch}
\end{align}
We explicitly denote the time dependence as a reminder that this is a rapidly-varied quantity. Of course, all receiver parameters are time-dependent to some extent. The two phase switch states are denoted by subscripts ``0'' for the zero-phase shift state and ``1'' for the $180\dg$ phase shift state. The amplitude of the transmission is given by $\sqrt{\alpha_{0,1}(t)}$ (ideally $=1$ in both states) and the phase shift is denoted by $\phi_{0,1}(t)$ (ideally $\phi_{0}=0, \phi_{1}=\pi$). A further consideration is that two phase switches may have different responses in supposedly equal states.

\subsubsection{Post-detection Gain}

Each detector diode output voltage, which is proportional to the input power to the detector diode, is transported, amplified, filtered and digitized by a different chain of electronics. These may have different gains. We model this by granting each detector diode a different responsivity $\alpha$ in Equations~\ref{eqn:rd:P_output_scattering} and \ref{eqn:Mueller_from_S}.

\section{Receiver Analysis} \label{sec:receiver_analysis}

At this point in the analysis we have used the procedure outlined in Section~\ref{sec:receiver_modelling} to calculate the receiver Mueller matrix and noise vector of the model described in Section~\ref{sec:C-BASS_receiver_model_and_errors}. The elements in the Mueller matrix and noise vector are complicated analytic expressions containing the variables described above. We will now simplify these analytic expressions to explore various aspects of the receiver performance.

\subsection{Ideal Receiver Behavior}

The receiver will behave in an ideal fashion when all the components are perfect. There is no amplitude or phase difference between the amplifiers: all voltage gains are equal to some gain factor $g$. The receiver Mueller matrix and noise vector, after dividing by the factor $\abs{g}^{2}$, are given by:
\begin{align}
\mathbf{M}_{\rm{rec}} = &
\begin{bmatrix}
1 & 0 & 0 & 0 \\
0 & 1 & 0 & 0 \\
0 & 0 & 1 & 0 \\
0 & 0 & 0 & 1
\end{bmatrix}\label{eqn:ideal_receiver:M} \\
\mathbf{N}_{\mathrm{rec}} = & 
\begin{bmatrix}
 -(T_{B}+T_{A})+DLT_{\rm ND}+D(1-L)T_{\mathrm{amb}} \\
0 \\
DLT_{\rm ND} + D(1-L)T_{\mathrm{amb}} \\
T_{B}-T_{A}
\end{bmatrix}. \label{eqn:ideal_receiver:N}
\end{align}

As expected, the ideal receiver Mueller matrix is the identity matrix. There is no leakage between Stokes parameters, and all data channels have identical gains. An offset term $-(T_{A}+T_{B})$ appears in the $r_{I}$ channel, indicating that what we measure is the difference between the sky total intensity and the reference load temperature. An unwanted offset of $T_{B}-T_{A}$ appears in the $r_{V}$ channel; this should, however, be zero if the reference loads are held at the same physical temperature.

An offset term $D\Big[ LT_{\rm ND} + (1-L)T_{\mathrm{amb}} \Big]$ appears in both the $r_{I}$ and $r_{U}$ channels. This is due to the calibration signal injection system, which is used to measure the instrument response by injecting a signal with known properties. This is discussed in more detail in the correlation receiver section \S\ref{sec:correlation_receiver_errors}.

\subsection{Role of Phase Switching} \label{sec:role_of_phase_switching}

Phase switching performs two roles in this receiver: it reduces the leakage of total intensity into the polarization channels and it modulates the slowly-varying sky signal at a high frequency. This modulation allows some undesired low-frequency signals that would otherwise contaminate the sky signal, such as low-frequency mains pickup, to be reduced by high-pass filtering prior to demodulation (or by low-pass filtering after demodulation). We will explore the role of phase switching in reducing leakage of total intensity to the polarization channels by considering a series of phase switching scenarios.

There are three pairs of phase switches in the receiver in Figure~\ref{fig:receiver_model}. The first pair, P14 and P15, switch the I1 output. The second pair, P17 and P18, switch the I2 output. The third pair, P22 and P23 switch the Q1, Q2, U1, and U2 outputs. The pairs are switched independently with orthogonal Walsh functions. We consider the general case where all phase switch states have both amplitude and phase errors, and the errors are different in the two phase switches. Since only the difference between the phase switches is important (any shared amplitude or phase error is mathematically degenerate with differing amplifier gains) we can make the 0 state for one of the phase switches an ideal scattering matrix. The scattering matrices for both phase switches in a pair in both phase switch states are shown in Table~\ref{tab:phase_switch_state_scattering_matrices}.

\begin{table}
\caption{The scattering matrices for phase switches 1 and 2 in both phase switch states. $\mathbf{S}_I$ is the 2x2 ideal transmission matrix. For ideal phase switches $\epsilon_{1}=\epsilon_{2}=\epsilon_{3}=\phi_{1}=\phi_{2}=\phi_{3}=0$.}
\label{tab:phase_switch_state_scattering_matrices}
\centering
\begin{tabular}{|c|c|c|}\hline
State $i$ & $\mathbf{S}_{PS1,i}$ & $\mathbf{S}_{PS2,i}$ \\ \hline
0 & $\mathbf{S}_I$ & $\sqrt{1-\epsilon_{1}}e^{-i\phi_{1}}\mathbf{S}_I$ \\
1 & $ \sqrt{1-\epsilon_{2}}e^{-i(\pi+\phi_{2})}\mathbf{S}_I$ & $\sqrt{1-\epsilon_{3}}e^{-i(\pi+\phi_{3})}\mathbf{S}_I$ \\ \hline
\end{tabular}
\end{table}

We assign the power detectors on outputs O1 to O12 responsivity coefficients $\alpha_{1}$ to $\alpha_{12}$ respectively. 
These represent the multiplication of differing detector diode sensitivities and differing post-detection gains. We make all the other components in the receiver perfect and make all gains equal to 1.

The action of phase switching can be revealed by looking at the first column of the instrument Mueller matrix. This encodes the contribution of the total intensity (Stokes $I$) to each of the four raw receiver data channels. For an ideal receiver only the first element is non-zero (see Equation~\ref{eqn:ideal_receiver:M}).

\subsubsection{No Phase Switching}

If the receiver had no phase switches -- i.e., assign the ideal transmission matrix to P14, P15, P17, P18, P22 and P23 -- the first column of the receiver Mueller matrix is:
\begin{align}
\begin{bmatrix}
 M_{II} \\ M_{QI} \\ M_{UI} \\ M_{VI}
\end{bmatrix} = 
\begin{bmatrix}
\frac{1}{2}(\alpha_{2}+\alpha_{12}) \\
\frac{1}{4}(\alpha_{6}-\alpha_{5}+\alpha_{9}-\alpha_{10}) \\
\frac{1}{4}(\alpha_{3}-\alpha_{4}+\alpha_{7}-\alpha_{8}) \\
\frac{1}{2}(\alpha_{2}-\alpha_{12})
\end{bmatrix}. \label{eqn:phase_switching:no_phase_switching}
\end{align}
Because we have set the amplitudes and phases of the gain chains to be equal, the load signal $T_{A}$ appears exclusively at O1 and has the coefficient $\alpha_{1}$ in the noise vector $\mathbf{N}_{\rm{rec}}$. Similarly, $T_{B}$ appears exclusively at output O11 and has the coefficient $\alpha_{11}$.

Without phase switching the total intensity leaks into the polarization signal channels \emph{if} there is an amplitude difference in the post-warm hybrid power detection hardware. This is very likely to be the case, so analogue correlation polarimeters must be phase switched.

\subsubsection{Asymmetric Phase Switching}

We now take the simplest approach to phase switching and ``jam'' one phase switch in each pair in a constant state, switching only the other. We call this asymmetric phase switching. We use the scattering matrices listed in Table~\ref{tab:phase_switch_state_scattering_matrices} for the phase switches and calculate the first column of the receiver Mueller matrix:
\begin{align}
\nonumber & M_{II} =  \frac{(\epsilon_{1}-\epsilon_{3})(\alpha_{1}-\alpha_{2}+\alpha_{11}-\alpha_{12})}{16} +\frac{1}{8}[\\
\nonumber & (\alpha_{1}+\alpha_{2}+\alpha_{11}+\alpha_{12})(\sqrt{1-\epsilon_{1}}\cos\phi_{1}+\sqrt{1-\epsilon_{3}}\cos\phi_{3}) ] \\
\nonumber &M_{QI} =  \frac{(\epsilon_{1}-\epsilon_{3})(\alpha_{5}-\alpha_{6}+\alpha_{10}-\alpha_{9})}{16}
\end{align}
\begin{align}
\nonumber &M_{UI} =  \frac{(\epsilon_{1}-\epsilon_{3})(\alpha_{4}-\alpha_{3}+\alpha_{8}-\alpha_{7})}{16} \\
\nonumber &M_{VI} =  \frac{(\epsilon_{1}-\epsilon_{3})(\alpha_{1}-\alpha_{2}-\alpha_{11}+\alpha_{12})}{16} +\frac{1}{8}[\\
 & (\alpha_{1}+\alpha_{2}-\alpha_{11}-\alpha_{12})(\sqrt{1-\epsilon_{1}}\cos\phi_{1}+\sqrt{1-\epsilon_{3}}\cos\phi_{3}) ] .
\end{align}

Introducing asymmetric phase switching has reduced the total intensity to polarization leakage -- the leakage of Stokes $I$ into $Q$ has been multiplied by a factor of $(\epsilon_{1}-\epsilon_{3})/4 \ll 1$ -- but not removed it entirely.

\subsubsection{Symmetric Phase Switching} \label{sec:phase_switching:symmetric_phase_switching}

The most general form of phase switching is when we switch both phase switches, spending an equal amount of time in each phase switch state for each data sample\footnote{This is sometimes called ``double demodulation''}. We average the states in which both are ``high'' or both are ``low'' to form a composite 0 state, and do the same for the states in which they are switched in an opposite sense to form a composite 1 state. The first column of the receiver Mueller matrix is now:
\begin{align}
\nonumber &M_{II} =  \frac{\alpha_{1}+\alpha_{2}+\alpha_{11}+\alpha_{12}}{16}\Big[\sqrt{1-\epsilon_{1}}\cos\phi_{1}+\sqrt{1-\epsilon_{3}}\cos\phi_{3} \\
\nonumber  &+ \sqrt{1-\epsilon_{2}}\Big(\sqrt{1-\epsilon_{1}}\cos(\phi_{1}-\phi_{2}) + \sqrt{1-\epsilon_{3}}\cos(\phi_{2}-\phi_{3})\Big) \Big] \\
\nonumber &M_{QI} =  0 \\
\nonumber &M_{UI} =  0 \\
\nonumber &M_{VI} =  \frac{\alpha_{1}+\alpha_{2}-\alpha_{11}-\alpha_{12}}{16}\Big[\sqrt{1-\epsilon_{1}}\cos\phi_{1}+\sqrt{1-\epsilon_{3}}\cos\phi_{3} \\
  &+ \sqrt{1-\epsilon_{2}}\Big(\sqrt{1-\epsilon_{1}}\cos(\phi_{1}-\phi_{2}) + \sqrt{1-\epsilon_{3}}\cos(\phi_{2}-\phi_{3})\Big) \Big]. \label{eqn:phase_switching:full_phase_switching}
\end{align}

Symmetric phase switching stops the leakage of total intensity into the polarization channels. Imperfections in the phase switches and differences in the power detection chains now merely manifest themselves as reductions in the gain of the receiver.

In summary, a major role of phase switching in a correlation polarimeter is to compensate for differences in the power detection chains. If uncorrected, these differences would result in a leakage of total intensity into the polarization channels. While asymmetric phase switching reduces this leakage it does not remove it entirely thanks to imperfections in the phase switches themselves -- symmetric phase switching is needed to do this. This result remains true if we reintroduce imperfections to all the post-OMT components. Stokes leakage due to imperfections in the OMT cannot be reduced by this type of phase switching as the Mueller matrices are cascaded -- see \S\ref{sec:OMTErrors}.

\subsection{Circularizing OMT Errors} \label{sec:OMTErrors}

The circularizing OMT is perhaps the most critical component in a polarimeter. Cleanly extracting orthogonal modes from a waveguide or free-space wave without some difference between the treatment of the modes, or leakage between them, is extremely difficult. Leakage between Stokes parameters caused by the OMT is impractical to correct in the instrument that follows; careful calibration of the receiver data is needed to correct for it.

In the model we use for the circularizing OMT scattering matrix (Equation~\ref{eqn:scattering:OMT}) there are three steps in the process of converting orthogonal linear modes in waveguide into circular polarization signals in cables.
\begin{enumerate}
\item Extraction of linear modes: The orthogonal linear probes might have differing transmission amplitudes ($\alpha_{x}$ and $\alpha_{y}$), and they may be misaligned ($\phi_{\perp}$).
\item Transmission of linear modes: The cables that connect the linear OMT to the circulariser may have different lengths, leading to a phase difference $\phi_{y}$. Note that any differences in the transmission amplitudes of these two cables are mathematically degenerate with $\alpha_{x}$ and $\alpha_{y}$, so are not separately parameterized.
\item Conversion to circular: The $90\dg$ hybrid that converts the linear polarizations into circular polarizations may have both amplitude ($\delta$) and phase ($\phi_{90}$) errors (see Equation~\ref{eqn:90hybrid_error_model} for the $90\dg$ hybrid scattering matrix).
\end{enumerate}

To illuminate how each of these steps affect the leakage between
Stokes parameters we write the Mueller matrix for the circularizing
OMT as the product of three Mueller matrices corresponding to the
steps:

\begin{align}
\mathbf{M}_{\rm{OMT}} = & \mathbf{M}_{90}\mathbf{M}_{\rm{trans}}\mathbf{M}_{\rm{lin}}. \label{eqn:omt_mueller}
\end{align}
We calculated these Mueller matrices by making all post-OMT components
in the receiver perfect. The Mueller matrix for the part of the
receiver that follows the circularizing OMT, $\mathbf{M}_{\rm{corr}}$, is then the identity matrix, and so the receiver Mueller matrix is equal to the circularizing OMT Mueller matrix.


\paragraph*{Extraction of linear modes:} The Mueller matrix for the linear step is:
\begin{tiny}
\begin{align}
\nonumber &\mathbf{M}_{\rm{lin}} =   \\
&\frac{1}{2}\begin{bmatrix}
\alpha_{x}+\alpha_{y} & \alpha_{x}-\alpha_{y}\cos2\phi_{\perp} & \alpha_{y}\sin2\phi_{\perp} & 0 \\
\alpha_{x}-\alpha_{y} & \alpha_{x}+\alpha_{y}\cos2\phi_{\perp} & -\alpha_{y}\sin2\phi_{\perp} & 0 \\
2\sqrt{\alpha_{x}\alpha_{y}}\sin\phi_{\perp} & 2\sqrt{\alpha_{x}\alpha_{y}}\sin\phi_{\perp} & 2\sqrt{\alpha_{x}\alpha_{y}}\cos\phi_{\perp} & 0 \\
0 & 0 & 0 & 2\sqrt{\alpha_{x}\alpha_{y}}\cos\phi_{\perp}
\end{bmatrix}.
\end{align}
\end{tiny}
This Mueller matrix leads to several insights. Leakage of Stokes $I$ into the measurement of $Q$ is proportional to the difference between the transmission amplitudes of the two probes. Leakage of Stokes $I$ into the measurement of $U$ is determined solely by the orientation error of the probes. Both these errors lead to leakage of the linear polarization parameters $Q$ and $U$ into the measurement of Stokes $I$, though this is not as grave a concern due to the small amplitude of the polarized signal in most astronomical applications.

\paragraph*{Transmission of linear modes:} The Mueller matrix for the transmission step is:
\begin{align}
&\mathbf{M}_{\rm{trans}} = 
\begin{bmatrix}
1 & 0 & 0 & 0 \\
0 & 1 & 0 & 0 \\
0 & 0 & \cos\phi_{y} & -\sin\phi_{y} \\
0 & 0 & \sin\phi_{y} & \cos\phi_{y}
\end{bmatrix}.
\end{align}
Phase or path length differences between the cables that connect the linear part of the OMT and the circularizing $90\dg$ hybrid lead to mixing between the $U$ and $V$ Stokes parameters via a rotation matrix, but does not lead to any leakage of total intensity into the linear polarization measurement.

\paragraph*{Circularizing:} The Mueller matrix for the circularizing step is:
\begin{tiny}
\begin{align}
\nonumber &\mathbf{M}_{90} = \\
&\begin{bmatrix}
1 & 0 & \sqrt{1-\delta^{2}}\sin\phi_{90} & 0 \\
0 & \sqrt{1-\delta^{2}}\cos\phi_{90} & 0 & -\delta \\
\sqrt{1-\delta^{2}}\sin\phi_{90} & 0 & 1 & 0 \\
0 & \delta & 0 & \sqrt{1-\delta^{2}}\cos\phi_{90}
\end{bmatrix}.
\end{align}
\end{tiny}
Errors in converting the linear polarization signals to a circular basis does not affect the measured Stokes $Q$ parameter (other than an amplitude change), but it can lead to leakage of Stokes $I$ into the measurement of Stokes $U$ if there is a non-zero phase error $\phi_{90}$ in the $90\dg$ hybrid.

\subsubsection{Systematic Offsets}

Errors in the circularizing OMT can also lead to systematic offsets in the data channels. The contribution of the circularizing OMT to the noise vector $\mathbf{N}_{\rm{rec}}$ (Equation~\ref{eqn:receiver_data_vector_definition}) is:
\begin{align}
\nonumber & \mathbf{N}_{\rm{OMT}} = T_{\rm{cryo}} \\
&
\begin{bmatrix}
1 - \alpha_y - \alpha_x \\
- \cos\phi_{90} \sqrt{1 - {\delta}^2}\, \left(\alpha_x - \alpha_y\right) \\
\frac{2\, \tan\!\left(\frac{\phi_{90}}{2}\right)\, \sqrt{1 - {\delta}^2}\, \left(\alpha_x - \alpha_y\right)}{{\tan\!\left(\frac{\phi_{90}}{2}\right)}^2 + 1} \\
\delta\, \left(\alpha_x - \alpha_y\right)
\end{bmatrix} \label{eqn:omt_noise_error}
\end{align}
where $T_{\rm{cryo}}$ is the physical temperature of the circularizing OMT. If the circularizing OMT is perfect this contribution becomes zero, as expected.

\subsection{Correlation Receiver Errors} \label{sec:correlation_receiver_errors}

In this section we assume that the circularizing OMT is ideal and investigate the action of an imperfect correlation architecture. The circularizing OMT and correlation Mueller matrices are combined multiplicitively and the noise offsets are combined additively (Equation~\ref{eqn:receiver_data_vector_definition}).

We assign complex voltage gains $g_{L1}$, $g_{L2}$, $g_{R1}$, $g_{R2}$, $g_{PL}$, and $g_{PR}$ to amplifiers P6, P7, P8, P9, P24, and P25 respectively. With perfect phase switches and equal detector sensitivities the Mueller matrix for the correlation receiver is given by:
\begin{align}
\mathbf{M}_{\rm{corr}} = 
& \begin{bmatrix}
 \alpha_{I}G_{\Sigma} & 0 & 0 & \alpha_{I}G_{\Delta}\\
0 & \alpha_{Q}G_{Pc} & \alpha_{Q}G_{Ps} & 0 \\
0 &-\alpha_{U}G_{Ps} & \alpha_{U}G_{Pc} & 0 \\
\alpha_{I}G_{\Delta} & 0 & 0 &  \alpha_{I}G_{\Sigma}
\end{bmatrix}. \label{eqn:full_receiver_Mueller}
\end{align}
The parameters used in Equations~\ref{eqn:full_receiver_Mueller} and \ref{eqn:full_receiver_Noise} are explicitly written out in Appendix~\ref{sec:appendix:full_receiver_model_definitions}. Note that the assumption of perfect phase switches and equal detector sensitivities is one of convenience for simplicity's sake. In \S\ref{sec:role_of_phase_switching} we showed that if we use symmetric phase switching these imperfections result in a small change to the amplitude of the Mueller matrix parameters, with no leakage of Stokes $I$ to $Q$ or $U$.

The vector of noise contributions to the data channels is:
\begin{small}
\begin{align}
\nonumber &\mathbf{N}_{\rm{corr}} = \\
&\begin{bmatrix}
-\alpha_{I}\Big\{ G_{L} T_{A} + G_{R} T_{B} 
- G_{\Sigma}D\Big(LT_{\rm ND}+(1-L)T_{\rm{amb}} \Big) \Big\}  \\
\alpha_{Q}G_{Ps}D[LT_{\rm ND}+(1-L)T_{\rm{amb}}] \\
\alpha_{U}G_{Pc}D[LT_{\rm ND}+(1-L)T_{\rm{amb}}] \\
 -\alpha_{I}\Big\{ G_{L} T_{A} - G_{R} T_{B}  
- G_{\Delta}D\Big(LT_{\rm ND}+(1-L)T_{\rm{amb}} \Big) \Big\}
\end{bmatrix}. \label{eqn:full_receiver_Noise}
\end{align}
\end{small}
Recall that the noise vector is not a measure of the variance of the measured signal: it is the systematic offset in each data channel due to noise produced in the instrument. 

\subsubsection{Interpretation of parameters:}
The Mueller matrix (Equation~\ref{eqn:full_receiver_Mueller}) and noise vector (Equation~\ref{eqn:full_receiver_Noise}) contain a number of composite parameters. $G_{\Sigma}$ is a phase-error-weighted average of the composite left section gain and the composite right section gain. If there are no phase errors in the $180\dg$ hybrids or the amplification chains it becomes $G_{\Sigma} = (\abs{g_{L1}}\abs{g_{L2}}+\abs{g_{R1}}\abs{g_{R2}})/2$. Phase differences between the gain chains change the contributions of $\abs{g_{L1}}\abs{g_{L2}}$ and $\abs{g_{R1}}\abs{g_{R2}}$ to $G_{\Sigma}$ as specified in Appendix~\ref{sec:appendix:full_receiver_model_definitions}. $G_{\Delta}$ is a phase-error-weighted difference of the composite gains. If there are no phase errors in the receiver it is given by $G_{\Delta}= (\abs{g_{L1}}\abs{g_{L2}}-\abs{g_{R1}}\abs{g_{R2}})/2$.

$\alpha_{I}$, $\alpha_{Q}$, and $\alpha_{U}$ are gain-reducing terms depending on hybrid errors -- both warm and cold $180\dg$ hybrids, and  warm $90\dg$ hybrids -- and are equal to unity in an ideal receiver. $G_{Pc}$ and $G_{Ps}$ are polarization gain terms and have a cosine/sine dependence for an ideal receiver, discussed in more detail in \S\ref{sec:measurement_of_linear_polarization}.

\subsubsection{Measurement of total intensity:}

The raw measurement of total intensity produced by the receiver is:
\begin{align}
\nonumber r_{I} =   k_{B}\alpha_{I}\Big\{ &\abs{g_{L1}}\abs{g_{L2}}(\alpha_{L1}T_{l}-\alpha_{L2}T_{A})  \\
\nonumber  + &\abs{g_{R1}}\abs{g_{R2}}(\alpha_{R1}T_{r}-\alpha_{R2}T_{B})  \\
  + &G_{\Delta}V+ N \Big\}
\end{align}
where the systematic offset is $N =  G_{\Sigma}D[LT_{\rm ND}+(1-L)T_{\rm{amb}}]$. $T_{l}$ and $T_{r}$ are the antenna temperatures of the left and right circular polarizations respectively (Equation~\ref{eqn:power_measured_in_each_polarization}), and are related to the Stokes total intensity $I$ by Equation~\ref{eqn:stokes_I_related_to_antenna_temperature}. There is a small leakage of circular polarization into the measurement of total intensity at the low level of $V/I = G_{\Delta}/G_{\Sigma}$. We have assumed a perfect circularizing OMT in this model, hence there is no leakage from $Q$ and $U$ into the measurement of $I$.

$\alpha_{L1}$ and $\alpha_{L2}$ differ only in the way they include $180\dg$ hybrid phase errors. If the $180\dg$ hybrids had no phase errors they would be equal. The power of the continuous comparison radiometer is immediately clear. A conventional radiometer measures $r_{c} \propto \abs{g}^{2}T_{l}$; a change in the receiver gain of $\Delta G$ causes a change in the output of $\Delta r_{c} = \Delta G T_{l}$. In a continuous comparison radiometer the change in the receiver output given the same gain change is $\Delta r_{I} = \Delta G (\alpha_{L1}T_{l}-\alpha_{L2}T_{A})$. The sensitivity to gain changes is reduced by a factor of $\frac{\alpha_{L1}T_{l}-\alpha_{L2}T_{A}}{T_{l}}$ compared to a conventional radiometer. We are \emph{completely} insensitive to receiver gain fluctuations if $\alpha_{L2}T_{A} = \alpha_{L1}T_{l}$.

\begin{figure}
 \centering
 \includegraphics[width=0.47\textwidth]{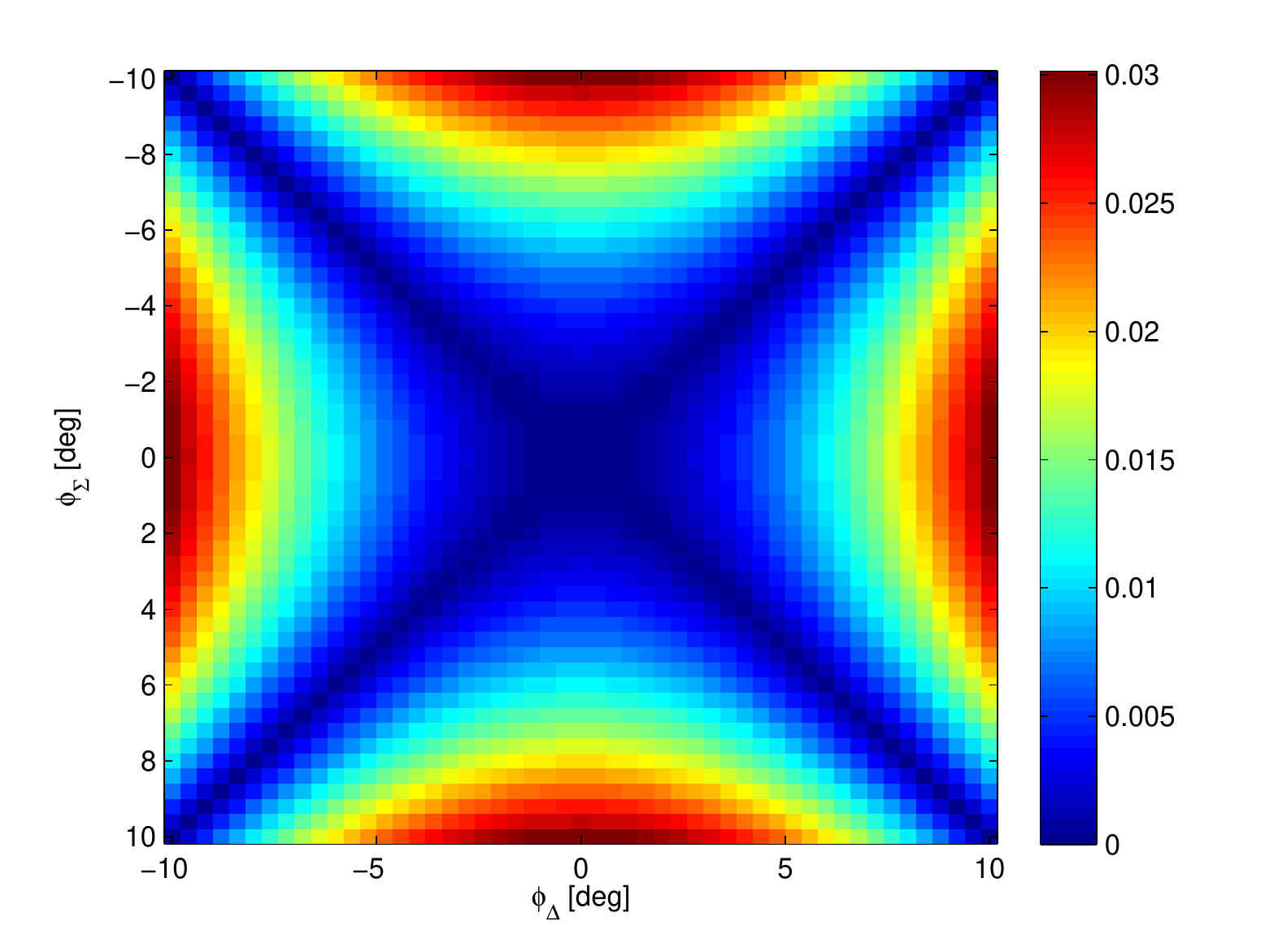}
\caption{The dependence on the $1/f$ noise in the $r_{I}$ channel on phase errors in the $180\dg$ hybrid, plotted in this heat map, is non-trivial. In general, phase errors increase the $1/f$ noise, but not if $\vert \phi_\Sigma \vert = \vert \phi_\Delta \vert$.
We assume perfect temperature balance, i.e., $T_{A}=T_{B}=T_{l,r}$, and that the warm and cold $180\dg$ hybrids are identical: $\phi_{\Delta}=\phi_{\Delta,w}=\phi_{\Delta,c}$ and $\phi_{\Sigma}=\phi_{\Sigma,w}=\phi_{\Sigma,c}$. We have also assumed no phase difference between the gain chains.}
 \label{fig:radiometer_1overf_dependence_on_hybrid_phase}
\end{figure}

The presence of $\alpha_{L1}$ and $\alpha_{L2}$ in the temperature difference term is not desirable, and depends only on $180^{\dg}$~phase errors. The output of a perfect continuous comparison radiometer should be proportional to the temperature difference between sky and load only. Figure~\ref{fig:radiometer_1overf_dependence_on_hybrid_phase} shows the fractional contribution of receiver gain fluctuations to the measured total intensity signal, as compared to a conventional radiometer, when the temperature balance is ideal ($T_{A}=T_{B}=T_{l,r}$). We have assumed that the warm and cold $180\dg$ hybrids are identical ($\phi_{\Delta}=\phi_{\Delta,w}=\phi_{\Delta,c}$ and $\phi_{\Sigma}=\phi_{\Sigma,w}=\phi_{\Sigma,c}$) and that the receiver gain chains are phase balanced. It's clear that the departure from ideal behaviour is very small: only 3\% variations are seen even with fairly large phase errors of $10\dg$. The 1/f noise is more likely to be dominated by differences between the load and sky temperature.

\begin{figure}
 \centering
 \includegraphics[width=0.47\textwidth]{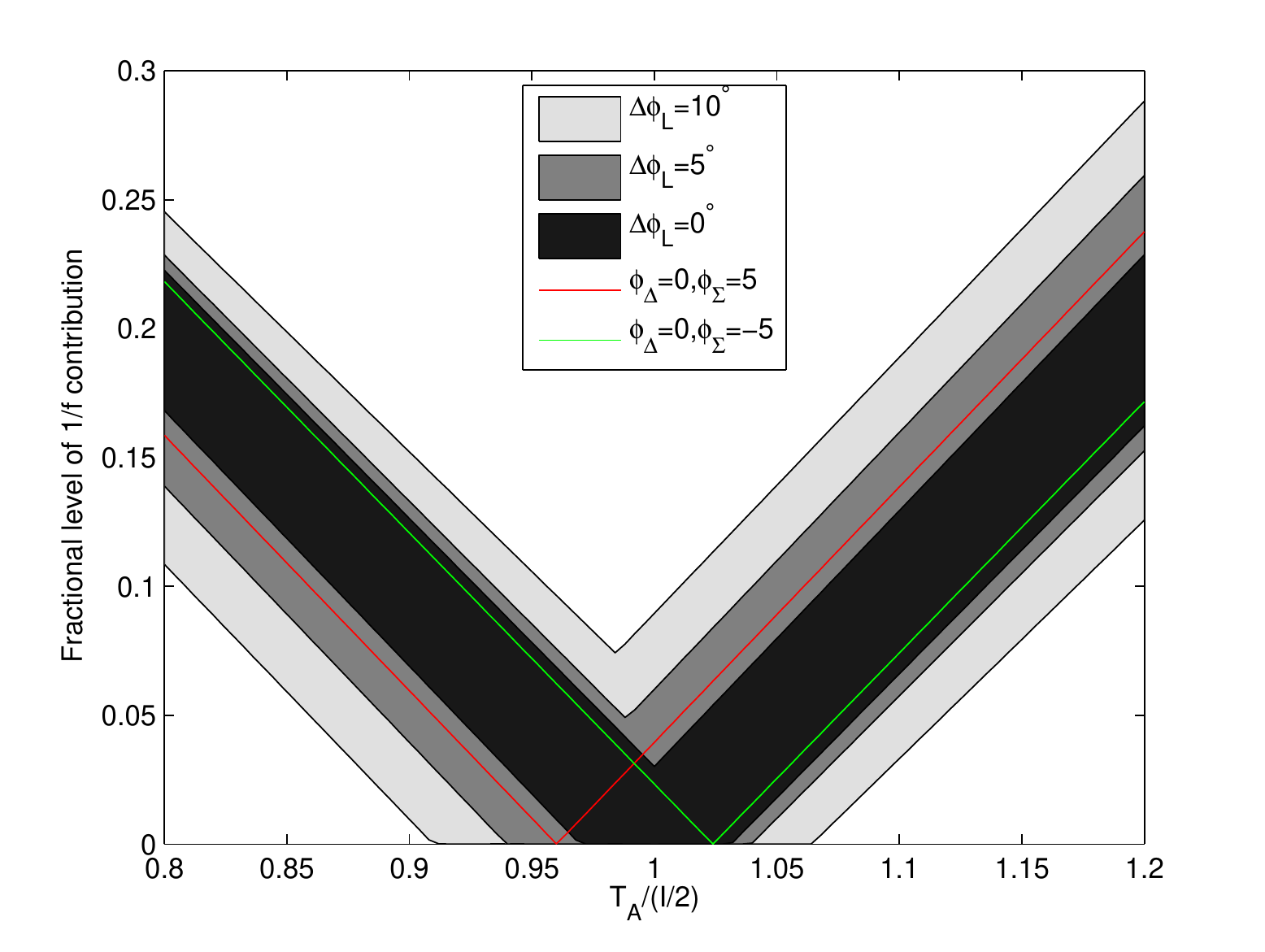}
\caption{Dependence of receiver 1/f noise in the $r_{I}$ channel -- as a fraction of what would be seen with a radiometer -- on temperature balance $\frac{T_{A}}{T_{l}}$ and gain chain phase error $\Delta\phi_{L}=\angle g_{L1}-\angle g_{L2}$. The shaded regions indicate the range of 1/f contribution found when the hybrid phase errors $\phi_{\Delta}$ and $\phi_{\Sigma}$ are independently varied between $-10\dg$ and $-10\dg$. Also plotted are two hybrid phase error cases for the $\Delta\phi_{L}=10\dg$ scenario. }
 \label{fig:radiometer_1overf_vary_temp_and_phase}
\end{figure}

The level of 1/f fluctuations seen in the total intensity data depends on 4 parameters: the gain chain phase difference $\Delta\phi_{L}=\angle g_{L1}-\angle g_{L2}$, the hybrid phase errors $\phi_{\Delta}$ and $\phi_{\Sigma}$, and the reference load to sky temperature ratio $\frac{T_{A}}{T_{l}}$. 
The effect of these parameters on the 1/f contribution is shown in Figure~\ref{fig:radiometer_1overf_vary_temp_and_phase}. We plot three gain chain phase difference scenarios where $\Delta\phi_{L}=0\dg$, $5\dg$ and $10\dg$. In each scenario we allow the hybrid phase errors $\phi_{\Delta}$ and $\phi_{\Sigma}$ to vary independently between $-10\dg$ and $10\dg$ and plot the range of 1/f contributions, as a fraction of the sky temperature $T_{l}$, seen in the $r_{I}$ channel versus $\frac{T_{A}}{T_{l}}$. Also plotted are two hybrid phase error cases ($\phi_{\Sigma}=5\dg,\phi_{\Delta}=0\dg$ and $\phi_{\Sigma}=-5\dg,\phi_{\Delta}=0\dg$) for the $\Delta\phi_{L}=10\dg$ scenario.

Figure~\ref{fig:radiometer_1overf_vary_temp_and_phase} underscores an important point: unequal phase in the two arms of a continuous comparison radiometer has a dramatic effect on the 1/f noise level, even for perfectly temperature balanced systems. Increasing the gain chain phase difference ``amplifies'' the effect of hybrid phase errors on the leakage: a larger level of 1/f noise might be found for the same hybrid performance. A gain amplitude difference between the arms, while it does affect the gain of the system, does not affect the relative level of 1/f noise.

\subsubsection{Measurement of linear polarization:} \label{sec:measurement_of_linear_polarization}
The power of correlation polarimetry is demonstrated by the first column of the Mueller matrix in Equation~\ref{eqn:full_receiver_Mueller}: there is no leakage of total intensity into the measurement of linear polarization. Put another way: even though we amplify the sky signal by very large, and unequal, factors of $10^{7}$ or more in a typical radio polarimeter this does not lead to \emph{any} leakage of total intensity into the raw polarization channels. 

The instrument mixes Stokes $Q$ and $U$ together with the matrix 
\begin{align}
 \mathbf{R}_{\rm{pol}} = 
\begin{bmatrix} 
\alpha_{Q}G_{Pc} & \alpha_{Q}G_{Ps}\\
-\alpha_{U}G_{Ps} & \alpha_{U}G_{Pc}
\end{bmatrix}.
\end{align}
Visualizing the variation of $\mathbf{R}_{\rm{pol}}$ with its 15 parameters is difficult. We instead discuss a few simplified scenarios.

\begin{figure}
 \centering
 \includegraphics[width=0.5\textwidth]{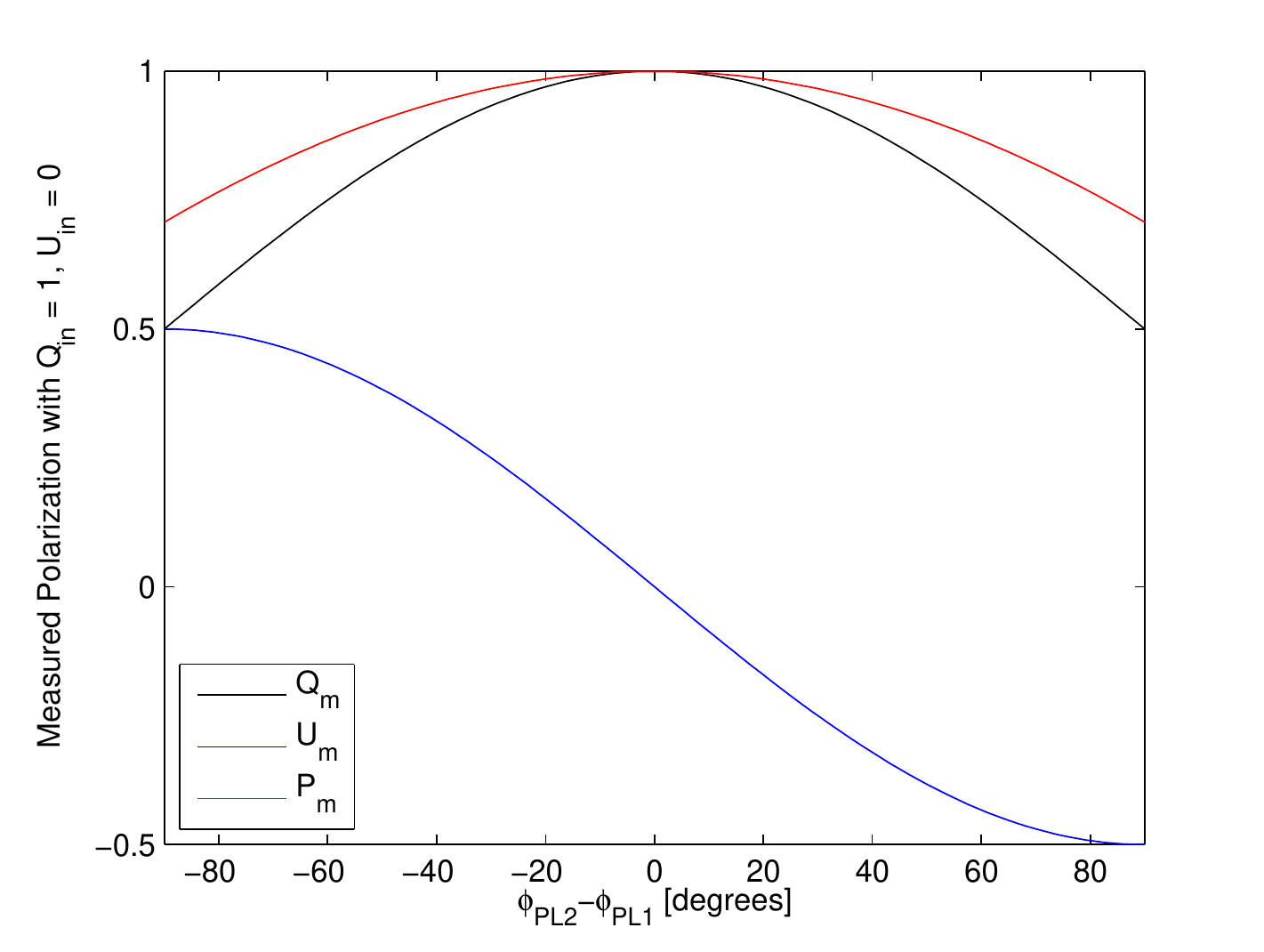}
 \caption{Dependence of the measured linear Stokes vector ($Q_{m},U_{m}$) and polarization amplitude $P_{m}=\sqrt{Q_{m}^{2}+U_{m}^{2}}$ on the phase difference $\phi_{PL2}-\phi_{PL1}$ between the two ``left'' gain chains. An input of $Q_{\rm in}=1,U_{\rm in}=0$ is assumed.}
 \label{fig:polarization_output_dependence_on_phase_err}
\end{figure}

\paragraph{Phase error type 1:} In an otherwise perfect polarimeter, what effect does a net phase difference between the ``left'' gain section and ``right'' gain section have? $\mathbf{R}_{\rm{pol}}$ simplifies to:
\begin{align}
\mathbf{R}_{\rm{pol}} = \abs{g}^{2}
\begin{bmatrix}
\cos\phi_{R} & -\sin\phi_{R} \\
\sin\phi_{R} & \cos\phi_{R}
\end{bmatrix}\label{eqn:rotation_matrix_pol}
\end{align}
where all amplifier gain amplitudes are equal to $\abs{g}$ and we have assumed $\phi_{PL1}=\phi_{PL2}=0$ and $\phi_{PR1}=\phi_{PR2}=\phi_{R}$. A net phase difference results in a rotation of the linear polarization vector.

\paragraph{Phase error type 2:} In an otherwise perfect polarimeter, what effect does a phase difference between the two ``left'' gain chains (or the two ``right'' gain chains) have? The rotation matrix simplifies to:
\begin{align}
\mathbf{R}_{\mathrm{pol}} = \frac{\abs{g}^2}{2}
\begin{bmatrix}
 \cos\phi_{PL1}+\cos\phi_{PL2} & \sin\phi_{PL1}+\sin\phi_{PL2} \\
 -\sin\phi_{PL1}+\sin\phi_{PL2} & \cos\phi_{PL1}+\cos\phi_{PL2}
\end{bmatrix}.
\end{align}
The effect of the phase difference $\phi_{PL2}-\phi_{PL1}$ on the measured Stokes parameters $Q_{m}$ and $U_{m}$, with an input $Q_{\rm in}=1,U_{\rm in}=0$, is shown in Figure~\ref{fig:polarization_output_dependence_on_phase_err}. This phase error rotates the pure $Q$ signal into $U$, and depolarizes the signal ($P_{m}=\sqrt{Q_{m}^{2}+U_{m}^{2}}$).

\paragraph{Gain difference:} In an otherwise perfect polarimeter, how does the measured polarization depend on the amplitude of the amplifier gains? In this case the rotation matrix simplifies to $\mathbf{R}_{\mathrm{pol}} =\frac{1}{4}(\abs{g_{PL1}}+\abs{g_{PL2}})(\abs{g_{PR1}}+\abs{g_{PR2}}) \mathbf{I}$, where $\mathbf{I}$ is the $2 \times 2$ identity matrix. The amplitudes of the amplifier gains affect both polarization channels equally, and simply change the gain of the system.

\paragraph{Hybrid errors:} In an otherwise perfect polarimeter, how do hybrid errors affect the measured linear polarization vector? The effect is the same as for the case just discussed: the rotation matrix becomes the identity matrix multiplied by a gain factor $\abs{g}^2\cos\phi_{90}\sqrt{1 - \delta_{90}^2}(\cos\phi_{\Delta}\sqrt{1 - \delta_{180}^2} + 1)/2$, where we have assumed that the cold and warm $180\dg$ hybrids are identical: $\delta_{180}=\delta_{180,c}=\delta_{180,w}$, and $\phi_{\Delta}=\phi_{\Delta,w}=\phi_{\Delta,c}$.

\subsubsection{Measurement of circular polarization:}
The circular polarization channel illustrates the disadvantage of measuring polarization by differencing two highly-amplified signals (Equation~\ref{eqn:CBASS_receiver_data_vector}). The leakage of total intensity into the $r_{V}$ channel is proportional the difference between the gains of the left and right channels, $\alpha_{L1}\abs{g_{L1}}\abs{g_{L2}}$ and $\alpha_{R1}\abs{g_{R1}}\abs{g_{R2}}$. The C-BASS receiver is not designed or intended to measure Stokes $V$ however; such a receiver would correlate $E_{x}(t)$ and $E_{y}(t)$.

\subsubsection{Systematic offsets:}
All the receiver channels have a systematic offset due to the noise diode coupling into the signal path, even when the noise diode is turned off. This is an unavoidable consequence of having a noise diode in the system. It is, however, small: the offset in the $r_{I}$ channel is equal to 0.29~K in units of antenna temperature ($D=-30$~dB, $L\simeq-7$~dB), and is constant in time.

\section{Testing the model} \label{sec:testing_the_model}

We tested the receiver model using two sets of data. These were a long stare at the North Celestial Pole (NCP), and observations of standard astronomical calibration sources.
The long stare at the NCP presented the receiver with an unchanging sky and a constant-thickness atmosphere. This is used to measure the stability of the instrument, as any fluctuations in the output of the receiver must be due to receiver gain changes or random fluctuations in the atmospheric emission. The stare was performed with phase switching turned on and with it turned off. We fired the noise diode to provide a calibration signal that we use to scale the data and rotate the $Q,U$ vector to the noise diode reference frame. Observations of polarization standards were used to characterize the leakage of the unpolarized emission $I$ into the $Q$ and $U$ channels.



\subsection{Scaling the data}


If we turn the noise diode on and off in rapid succession and measure the difference in the receiver channels between the on and off states we get:
\begin{align}
\Delta \mathbf{r} =  
\begin{bmatrix}
\alpha_{I}G_{\Sigma}DL\Delta T_{\rm ND} \\
\alpha_{Q}G_{Ps}DL\Delta T_{\rm ND} \\
\alpha_{U}G_{Pc}DL\Delta T_{\rm ND} \\
\alpha_{I}G_{\Delta}DL\Delta T_{\rm ND}
\end{bmatrix}.
\end{align}

If the receiver were ideal, the noise diode would appear as a pure-$U$ source (Equation~\ref{eqn:ideal_receiver:N}). We can use this to measure the polarization angle rotation introduced by the receiver ($\phi_R$ in Equation~\ref{eqn:rotation_matrix_pol}) and apply the corrective rotation matrix that rotates the polarization vector back to the noise diode reference frame. 

\begin{figure}
 \centering
 \includegraphics[width=0.47\textwidth]{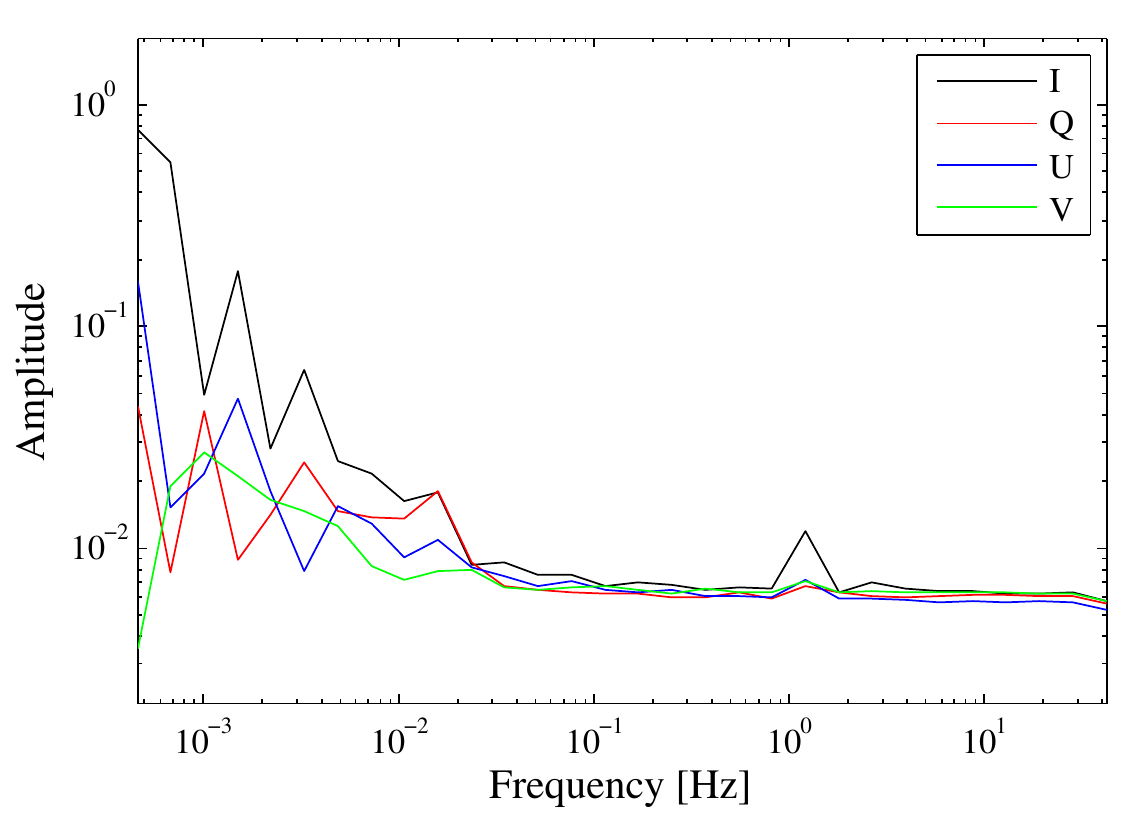}
\caption{Power spectrum of NCP stare data 
with phase switching turned on. The $I$ spectrum shows $1/f$ fluctuations at long timescales due (primarily) to the atmosphere. This is seen leaking into the $Q$, $U$, and $V$ channels, which should show no such intrinsic fluctuations due to atmospheric emission. The strong signal at $1.2\,$Hz is contamination from the cryogenic refridgerator and is removed from the data by the data reduction pipeline.}
 \label{fig:power_spectrum_I_and_P}
\end{figure}

\subsection{Phase switching}

In \S\ref{sec:role_of_phase_switching} we predicted that symmetric phase switching should reduce the leakage of total intensity in to the linear polarization channels. The power spectra of the phase-switched data are shown in Figure~\ref{fig:power_spectrum_I_and_P}. The emission from the atmosphere has a characteristic $1/f$ power spectrum, as seen clearly in the $I$ data. 

\begin{figure}
 \centering
 \includegraphics[width=0.47\textwidth]{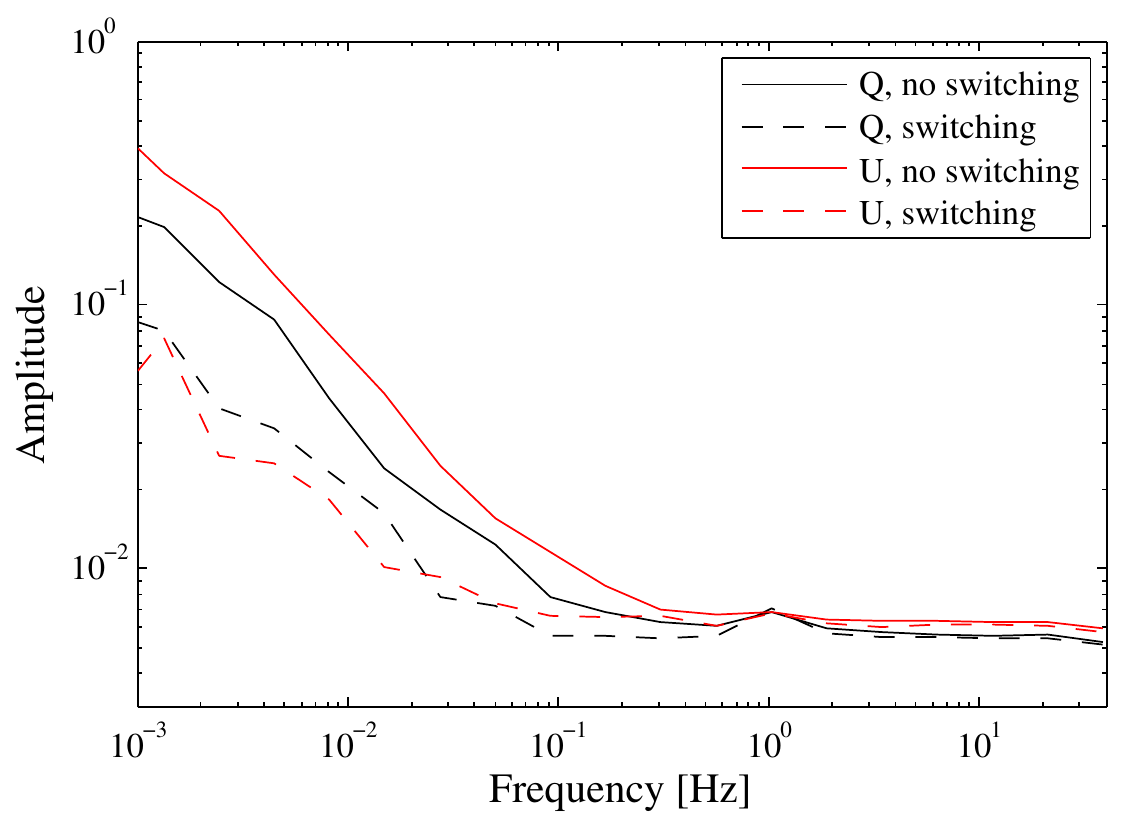}
\caption{The effect of phase switching on the receiver performance. The telescope was pointed at an unchanging sky (the North celestial pole) and data taken with symmetric phase switching on, and with no phase switching. The level of $1/f$ noise (which is expected from atmospheric fluctuations) is substantially reduced when phase switching is turned on, as expected by comparing Equations \ref{eqn:phase_switching:no_phase_switching} and \ref{eqn:phase_switching:full_phase_switching}. The remaining $1/f$ noise is due to $I$ to $Q,U$ leakage in the circularizing OMT and correlated amplifier noise.}
 \label{fig:phase_switching_effect_on_power_spectrum}
\end{figure}

These fluctuations in the atmospheric emission are inherently unpolarized, so their appearance in the $Q$ and $U$ channels are due to leakage from $I$ to $Q,U$. 
We can use the presence of this signal in the $Q,U$ channels to test the effect of phase switching. This is shown in Figure~\ref{fig:phase_switching_effect_on_power_spectrum}. The level of $1/f$ noise in the $Q$ and $U$ channels is substantially reduced by turning symmetric phase switching on. This is because phase switching removes the leakage of $I$ to $Q,U$ caused by differences in the post-second hybrid hardware, as shown in \S\ref{sec:phase_switching:symmetric_phase_switching}.

The remaining $1/f$ fluctuations in the $Q$ and $U$ channels are due to leakage terms in the circularizing OMT and a second effect that we have not modelled, namely correlated noise from the amplifiers. The analysis assumes that noise from different amplifiers is not correlated, but common temperature fluctuations can induce such a correlation. 

\subsection{OMT leakage: $I$ to $Q,U$}

We used observations of polarized and unpolarized standard sources to measure the $I$ to $Q,U$ leakage terms due to the circularizing OMT. 
We know from the modelling in \S\ref{sec:correlation_receiver_errors} that once symmetric phase switching is performed the only instrumental source of leakage into $Q$ and $U$ is due to the circularizing OMT. Other effects, such as cross-polar terms from the telescope optics, may be included in the analysis if desired, but are not addressed here.

An alternative means of estimating the $I$ to $Q,U$ leakage terms is to perform a skydip: by scanning the telescope from a low elevation to a high elevation we can vary the unpolarized atmospheric emission, hence change the Stokes $I$ input supposedly without changing the $Q$, $U$, or $V$ inputs. However, several significant sources of contamination affect this measurement: the polarized ground emission varies, particularly at low elevations; the telescope might scan through polarized galatic emission; and bright sources such as the sun or moon might cause a varying polarized input if they are present in the telescope sidelobes. For these reasons we instead used early on-axis observations of astronomical calibrators (Muchovej et al. in preparation) to measure the OMT leakage terms, and estimate the leakage of $I$ to $Q$ to be $m_{QI} = 0.5\,$\%, and the leakage from $I$ to $U$ to be $m_{UI} = -0.5\,$\%.

\subsubsection{Model fit}

If we calculate $\mathbf{M}_{\rm OMT}$ according to Equation~\ref{eqn:omt_mueller} and divide the leakage terms by the $II$ term, we get:
\begin{align}
\nonumber m_{QI} = & \frac{\cos\!\phi_{90}\, \sqrt{1 - \delta^2}\, \left(\frac{\alpha_{x}}{2} - \frac{\alpha_{y}}{2}\right) - \delta\, \sin\!\phi_{\perp}\, \sin\!\phi_{y}\, \sqrt{\alpha_{x}\, \alpha_{y}}}{\frac{\alpha_{x}}{2} + \frac{\alpha_{y}}{2} + \cos\!\phi_{y}\, \sin\!\phi_{90}\, \sin\!\phi_{\perp}\, \sqrt{\alpha_{x}\, \alpha_{y}}\, \sqrt{1 - \delta^2}} \\
m_{UI} = &\frac{\sin\!\phi_{90}\, \sqrt{1 - \delta^2}\, \left(\frac{\alpha_{x}}{2} + \frac{\alpha_{y}}{2}\right) + \cos\!\phi_{y}\, \sin\!\phi_{\perp}\, \sqrt{\alpha_{x}\, \alpha_{y}}}{\frac{\alpha_{x}}{2} + \frac{\alpha_{y}}{2} + \cos\!\phi_{y}\, \sin\!\phi_{90}\, \sin\!\phi_{\perp}\, \sqrt{\alpha_{x}\, \alpha_{y}}\, \sqrt{1 - \delta^2}}. \label{eqn:leakage_terms_to_Q_U}
\end{align}

We fit the expressions in Equation~\ref{eqn:leakage_terms_to_Q_U} to the measured leakage. This is clearly an ill-conditioned problem, as we have 6 variables and only 2 data points. However, we can produce an excellent fit using only 2 of the 6 variables as shown in Table~\ref{tab:OMT_model_parameters_fit}. This model fit implies that the difference in transmission of the power in the two linear modes $\alpha_x - \alpha_y$ is very small, $\sim 0.9$\%, and the effective error in the orientation of the OMT probes is $0.\!\dg3$. These are reasonable values.

\begin{table}
\caption{The best-fit model to the circularizing OMT leakage terms. Only 2 of the 6 model parameters, $\alpha_y$ and $\phi_{\perp}$, are needed to fit to the data.}
\label{tab:OMT_model_parameters_fit}
\centering
\begin{tabular}{|c|c|c|c|c|c|}\hline
$\alpha_x$ & $\alpha_{y}$ & $\phi_{\perp}$ & $\phi_{y}$ & $\phi_{90}$ & $\delta$ \\ \hline
1 & 0.991 & $-0.\!\dg3$ & $0\dg$ & $0\dg$ & 0 \\ \hline
\end{tabular}
\end{table}

\subsection{Correlation receiver leakage: $I$ to $V$}

The circularizing OMT-derived leakage to the Stokes $V$ channel is:
\begin{align}
m_{VI} = \frac{\delta(\alpha_{x} - \alpha_{y}) + 2\, \cos\!\phi_{90}\, \sin\!\phi_{\perp}\, \sin\!\phi_{y}\, \sqrt{\alpha_{x}\, \alpha_{y}}\, \sqrt{1 - \delta^2}}{\alpha_{x} + \alpha_{y} + 2\, \cos\!\phi_{y}\, \sin\!\phi_{90}\, \sin\!\phi_{\perp}\, \sqrt{\alpha_{x}\, \alpha_{y}}\, \sqrt{1 - \delta^2}}.
\end{align}

We measure a full-receiver leakage term from $I$ to $V$ of $m_{VI} = -1.4$\%. The values of $\alpha_x,\alpha_y,\delta$ in the OMT error model required to produce the measured leakage are infeasibly large. It is more likely that this $I$ to $V$ leakage comes from the correlation receiver, as modelled in Equation~\ref{eqn:full_receiver_Mueller}. 

This implies that the difference of the gains of the left and right channels is $G_\Delta/G_\Sigma = -0.014$. The data used to produce the fit have been scaled by the amplitude of the noise diode, so this gain difference has been removed. This means that the noise diode signal is not being injected with equal amplitude into the left and right channels. The difference in the injected powers of $1.4$\% is very reasonable: the noise diode signal is split and carried to the injection points by long cables whose transmission could easily differ by this amount. It is also possible that the directional couplers used to inject the noise diode signal into the signal path could differ by this much in their coupling constant.
%
%
%

\section{Conclusions} \label{sec:conclusions}

\begin{itemize}
\item \textbf{Powerful modelling approach:} We have described a modelling approach that enables us to produce a full signal and noise description of a receiver. By describing each component of the receiver with a scattering matrix we can take advantage of a mathematical framework that naturally includes both reflections from components and the noise produced by them. We applied this to a novel receiver, the northern C-BASS instrument, which is a hybrid of two commonly used architectures: a continous comparison radiometer for measuring total intensity, and a correlation polarimeter for measuring linear polarization. We are able to draw a number of conclusions about the behaviour of the C-BASS receiver architecture that are more broadly applicable to radio receivers.
\item \textbf{Phase Switching:} A crucial role of phase switching as it is implemented in the C-BASS receiver is to reduce leakage of Stokes $I$ into the measurement of linear polarization. We have described exactly how phase switching affects a correlation polarimeter in three modes. If no phase switching is implemented, gain differences between the power detection chains leads to leakage. Asymmetric phase switching reduces the level of this leakage, but does not remove it entirely due to imperfections in the phase switches. Symmetric phase switching cancels the effects of phase switch imperfections and removes the $I\rightarrow Q,U$ leakage entirely from the raw data.
\item \textbf{Correlation Polarimetry:} The full receiver model reveals that the post-OMT receiver, if symmetrically phase switched, causes \emph{no} leakage of total intensity into the polarization channels. The model is an accurate representation of the real receiver, though it is possible that effects not modelled here may cause some small level of leakage. This underscores the power of correlation polarimetry.
\item \textbf{Mode Separation:} The circularizing OMT, which separates orthogonal electric field modes of the incident electromagnetic wave, is the most crucial component in a correlation polarimeter. We describe exactly how errors in the extraction of linear modes from waveguide and in the conversion of linear signals to a circular basis both cause leakage of Stokes $I$ into $Q$ and $U$. 
\end{itemize}




The power of this approach to receiver modelling is clear: it allows the final instrument response to be evaluated in terms of individual receiver component imperfections. This enables us to identify which components are most critical for good instrument performance, and guides the data calibration process by producing an exact description of the leakage between Stokes parameters and the offsets due to component imperfections and noise. It enables us to gain a deep intuitive understanding of the operation, strengths and weaknesses of complicated receiver architectures.


\section*{Acknowledgements}

The C-BASS project is a collaboration between Caltech/JPL in the US, Oxford and Manchester Universities in the UK, and Rhodes University and the Hartebeesthoek Radio Astronomy Observatory in South Africa. It is funded by the NSF (AST-0607857, AST-1010024, and AST-1212217), the University of Oxford, the Royal Society, and the other participating institutions.
We thank Russ Keeney for technical support of the C-BASS project at OVRO. 
We thank the Xilinx University Programme for their donation of FPGAs to this project. 
OGK acknowledges the support of a Dorothy Hodgkin Award in funding his studies while a student at Oxford, and the support of a W.M. Keck Institute for Space Studies Postdoctoral Fellowship at Caltech. 
ACT acknowledges support from a Royal Society Dorothy Hodgkin Fellowship. 
CC acknowledges the support of the Commonwealth Scholarship, Square Kilometer Array South Africa and Hertford College.

\bibliographystyle{mn2e}
\bibliography{papersexport}

\clearpage
\appendix

\section{Stokes Parameters In Temperature Units} \label{sec:stokes_in_temp_units}

We would like to rewrite the Stokes parameters in terms of antenna temperate -- a more natural unit for a radio astronomy receiver.

We adopt the convention that $S_{0}, S_{1}, S_{2}, S_{3}$ refer to the Stokes parameters in units of brightness ($\rm{W}\rm{Hz}^{-1}\rm{m}^{-2}\rm{sr}^{-1}$), while $I,Q,U,V$ are the Stokes parameters in units of antenna temperature (K).

In the Rayleigh-Jeans limit, the relationship between the brightness of the sky  ($B_{\nu}$), its brightness temperature ($T_{b}$), and the Stokes parameters ($S_{0}$ to $S_{3}$ in brightness units) is given by:
\begin{align}
\nonumber B_{\nu} [\rm{W}/\rm{Hz}/\rm{m}^{2}/\rm{sr}] = & \frac{2 k_{B}T_{b}}{\lambda^{2}} \\
 = & \frac{1}{2}\left[ (S_{0}+S_{3}) + (S_{0}-S_{3}) \right].
\end{align}

The amount of power per unit bandwidth received by an antenna in each polarization is (assuming a circular basis):
\begin{align}
\nonumber P_{l} = & \frac{1}{2}A_{e} \int \int [S_{0}(\vartheta,\varphi)+S_{3}(\vartheta,\varphi)]P_{n}(\vartheta,\varphi)d\Omega \\
\nonumber = & k_{B}T_{l} \\
\nonumber P_{r} = & \frac{1}{2}A_{e} \int \int [S_{0}(\vartheta,\varphi)-S_{3}(\vartheta,\varphi)]P_{n}(\vartheta,\varphi)d\Omega \\
 & = k_{B}T_{r} \label{eqn:power_measured_in_each_polarization}
\end{align}
where $A_{e}$ is the effective aperture of the telescope, $P_{n}(\vartheta,\varphi)$ is the normalized power pattern of the antenna, and $T_{l}$ and $T_{r}$ are the antenna temperatures of each mode. For an unpolarized sky, $T_{l}=T_{r}$.

From the Nyquist theorem we know that the voltage $E(t)$ produced by a resistor with resistance $R$ and physical temperature $T$ has a variance:
\begin{align}
 \left\langle \abs{E(t)}^{2} \right\rangle = 4Rk_{B}T \label{eqn:nyquist_theorem}
\end{align}

We can now relate the Stokes total intensity in units of antenna temperature to the brightness:
\begin{align}
\nonumber k_{B} I = & k_{B}(T_{l}+T_{r}) \\
 = & A_{e}\int \int B_{\nu} (\vartheta,\varphi) P_{n}(\vartheta,\varphi)d\Omega. \label{eqn:stokes_I_related_to_antenna_temperature}
\end{align}

\clearpage
\newpage

\section{Full Receiver Model} \label{sec:appendix:full_receiver_model_definitions}

In this analysis we allow the cold and warm $180\dg$ hybrids to have different errors, even though the same type of component is used in both instances. The [cold/warm] hybrid errors are given by $\delta_{180,[c/w]}$, $\phi_{\Sigma,[c/w]}$, and $\phi_{\Delta,[c/w]}$. 

\begin{align*}
& \rm{Parameters\ used\ in\ Stokes\ } I \rm{\ and\ } V: \\
G_{L} = & \alpha_{L2} \abs{g_{L1}}\abs{g_{L2}} \\
G_{R} = & \alpha_{R2} \abs{g_{R1}}\abs{g_{R2}} \\
G_{\Sigma} = & \frac{ \alpha_{L1}\abs{g_{L1}}\abs{g_{L2}} + \alpha_{R1}\abs{g_{R1}}\abs{g_{R2}} }{2} \\
G_{\Delta} = & \frac{ \alpha_{L1}\abs{g_{L1}}\abs{g_{L2}} - \alpha_{R1}\abs{g_{R1}}\abs{g_{R2}} }{2} \\
\alpha_{I} = & \sqrt{(1-\delta_{180,c}^{2})(1-\delta_{180,w}^{2})} \\
\alpha_{[L,R]1} = & \frac{1}{2}\Big[ \cos(\angle g_{[L,R]1}-\angle g_{[L,R]2} +\phi_{\Delta,c}-\phi_{\Sigma,w}) 
  +\cos(\angle g_{[L,R]1}-\angle g_{[L,R]2} +\phi_{\Delta,c}+\phi_{\Delta,w}) \Big] \\
\alpha_{[L,R]2} = & \frac{1}{2}\Big[ \cos(\angle g_{[L,R]1}-\angle g_{[L,R]2} +\phi_{\Delta,w}-\phi_{\Sigma,c}) 
  +\cos(\angle g_{[L,R]1}-\angle g_{[L,R]2} -\phi_{\Sigma,c}-\phi_{\Sigma,w}) \Big]  \\
& \rm{Parameters\ used\ in\ Stokes\ } Q \rm{\ and\ } U: \\
G_{Pc} = & \frac{1}{4}\Bigg\{\sqrt{1-\delta_{180,c}^{2}}
  \Big[ \abs{g_{PL1}}\abs{g_{PR2}}\cos(\phi_{PL1}-\phi_{PR2}+\phi_{\Delta,c}) 
 + \abs{g_{PL2}}\abs{g_{PR1}}\cos(\phi_{PR1}-\phi_{PL2}+\phi_{\Delta,c}) \Big] \\
& + (1-\delta_{180,c})\abs{g_{PL1}}\abs{g_{PR1}}\cos(\phi_{PL1}-\phi_{PR1})  
 + (1+\delta_{180,c})\abs{g_{PL2}}\abs{g_{PR2}}\cos(\phi_{PL2}-\phi_{PR2}) \Bigg\}\\
G_{Ps} = & \frac{1}{4}\Bigg\{\sqrt{1-\delta_{180,c}^{2}} 
 \Big[ \abs{g_{PL1}}\abs{g_{PR2}}\sin(\phi_{PL1}-\phi_{PR2}+\phi_{\Delta,c}) 
 - \abs{g_{PL2}}\abs{g_{PR1}}\sin(\phi_{PR1}-\phi_{PL2}+\phi_{\Delta,c}) \Big] \\
& + (1-\delta_{180,c})\abs{g_{PL1}}\abs{g_{PR1}}\sin(\phi_{PL1}-\phi_{PR1})  
 + (1+\delta_{180,c})\abs{g_{PL2}}\abs{g_{PR2}}\sin(\phi_{PL2}-\phi_{PR2})\Bigg\}  \\
\alpha_{Q} = & \cos\phi_{90}\sqrt{1-\delta_{90}^{2}} \\
\alpha_{U} = & \frac{\cos\phi_{\Delta,w}+\cos\phi_{\Sigma,w}}{2}\sqrt{1-\delta_{180,w}^{2}} \\
& \rm{Where:}  \\
g_{PL}g_{L1} = & \abs{g_{PL1}}e^{-i\phi_{PL1}} \\
g_{PL}g_{L2} = & \abs{g_{PL2}}e^{-i\phi_{PL2}} \\
g_{PR}g_{R1} = & \abs{g_{PR1}}e^{-i\phi_{PR1}} \\
g_{PR}g_{R2} = & \abs{g_{PR2}}e^{-i\phi_{PR2}} 
\end{align*}

\label{lastpage}

\end{document}